\documentclass[aip,rsi,reprint,superscriptaddress,floatfix]{revtex4-1}
\usepackage{graphicx, amsmath}
\usepackage{natbib} 
\usepackage{color}
\usepackage{bm}

\definecolor{dgreen}{RGB}{0,128,0}

\begin{document}

\title{A multi-axis confocal rheoscope for studying shear flow of structured fluids}

\author{Neil Y.C. Lin}
\homepage[]{N. Y. Lin and J. H. McCoy contributed equally to this work.}
\affiliation{Department of Physics, Cornell University, Ithaca, NY 14853}

\author{Jonathan H. McCoy}
\homepage[]{N. Y. Lin and J. H. McCoy contributed equally to this work.}
\affiliation{Department of Physics and Astronomy, Colby College, Waterville, ME 04901}

\author{Xiang Cheng}
\affiliation{Department of Physics, Cornell University, Ithaca, NY 14853}

\author{Brian Leahy}
\affiliation{Department of Physics, Cornell University, Ithaca, NY 14853}

\author{Jacob N. Israelachvili}
\affiliation{Department of Chemical Engineering, and Materials Research Laboratory, University of California, Santa Barbara, CA 93106}

\author{Itai Cohen}
\affiliation{Department of Physics, Cornell University, Ithaca, NY 14853}

\date{\today} 

\begin{abstract}
We present a new design for a confocal rheoscope that enables uniform uniaxial or biaxial shear. The design consists of two precisely-positioned parallel plates with a gap that can be adjusted down to $2\pm$0.1 $\mu$m, allowing for the exploration of confinement effects. 
By using our shear cell in conjunction with a biaxial force measurement device and a high-speed confocal microscope, we are able to measure the real-time biaxial stress while simultaneously imaging the material 3D structure. We illustrate the importance of the instrument capabilities by discussing the applications of this instrument in current and future research topics in colloidal suspensions.
\end{abstract}

\maketitle

\section{Introduction}

In many systems driven far from thermodynamic equilibrium, deformation plays a decisive role.  Structured fluids in particular exhibit a rich array of non-equilibrium flow phenomena that impact natural and industrial processes alike \cite{pincus2004structured, Larson1998}.  Structured fluids have microscopic components, such as suspended particles, polymers, or micelles, whose distribution and dynamics can strongly affect bulk properties such as viscosity.  Thus, changes in microstructure arising from imposed flow conditions can have significant consequences.  Understanding the cooperative dynamics of entangled polymers is important for the processing of thermoplastics \cite{de1979scaling}, for example, and the interplay between platelet aggregation and blood flow is important in blood clotting \cite{born1963aggregation}.  Likewise, the non-Newtonian rheology of structured fluids can be exploited in the design of useful materials, such as the shear-thinning properties of paint or toothpaste or the possible use of shear-thickening to engineer flexible body armor \cite{wagner2009shear, mewis2012colloidal}.  The microscopic underpinnings of such phenomena and the range of microscopic behaviors observed in structured fluids  more generally have been the subject of considerable discussion.  The multi-scale, many-body character of these problems continues to challenge our physical understanding of systems far from equilibrium.

Imposing flow conditions in which the velocity gradient tensor takes a relatively simple form is a classic strategy for investigating structured fluids \cite{Larson1998, Macosko}.  A strain-controlled measurement performed on a commercial rheometer, for example, provides precise data on how shear stress varies with applied shear rates.  Alternatively, these instruments are capable of stress-controlled measurements and oscillatory measurements, as well as impulsive measurements in which transients can be explored.  These bulk rheological techniques can be used to characterize viscous and elastic responses, including nonlinear behaviors such as shear thinning and thickening, thixotropy, and yielding commonly observed in structured fluids \cite{Larson1998, Macosko, mewis2012colloidal}.  
In addition, by modifying the motor of a commercial rheometer and using a cylindrical Couette geometry, it is possible to superimpose a small-amplitude oscillatory motion orthogonal to a primary shear flow \cite{Vermant1997, vermant1998orthogonal, walker2000orthogonal, farage2012three}.  This modification enables biaxial flow measurements and can be used to probe force signatures arising from anisotropies in flow-induced structures.

While rheology does supply a great deal of useful information, deeper understanding generally requires {\it in situ} measurement techniques that provide access to the fluid's microscopic degrees of freedom, {\it e.g.}, polymer orientation, flow velocity field, or particle positions.  Scattering techniques provide powerful tools for probing microscopic structures.  Small angle x-ray and neutron scattering (SAXS and SANS), for example, are able to resolve the nanoscale structures and dynamics associated with polymeric systems.  Dynamic light scattering (DLS) and other light scattering techniques \cite{van1991dynamic, kroon1996dynamic, guo2000spatial, van1993dynamic}, by comparison, are less expensive, and easier to incorporate into a table-top experiment but are limited to structures no smaller than the wavelengths associated with visible light.   When combined with conventional rheology these techniques  offer valuable insights into the microscopic origins of observed flow behaviors in a variety of systems \cite{bender1996reversible, pileni1993reverse, schultz2000structural, somani2002shear}.  However, the structural information provided by scattering techniques is averaged over a large volume of sample, so point defects \cite{schall2006visualizing, schall2004visualization}, grain boundaries \cite{nagamanasa2011confined}, shear bands \cite{Cohen2006, moller2008shear, Besseling2010}, and other heterogeneities \cite{kegel2000direct} can be difficult to resolve.  Moreover, real-space structures can only be extracted from scattering data through Fourier analysis and, due to missing phase information, this process is not always straightforward.  

Such limitations can be overcome by a variety of real space imaging techniques\cite{nagamanasa2011confined, ou2004laminar, pogodina2001rheology}. More recently, for samples with structures larger than the wavelength of light, confocal microscopy has been used to measure a material's microstructure under shear. Using this technique, individual structures of interest can be followed in real time or, by scanning through the sample, the full three-dimensional structure can be mapped out, allowing for accurate reconstruction of flow profiles using particle velocimetry and detailed measurements of a material's dynamic microstructure \cite{Besseling2007,  Dutta2013, Petekidis2002, Derks2004, Wu2007, gao2010direct}. 

Two types of experimental designs combining confocal microscopy with precise flow control have been reported.  The major distinction between these two types is the geometry of the flow.  The first group of designs uses counter-rotating surfaces to drive torsional flows with circular streamlines \cite{Derks2004, Derks2009, Ballesta2008, Besseling2009, Besseling2010, Schmoller2010, Dutta2013}.  Often, one of these surfaces is fixed in the laboratory frame.  Independently rotating both surfaces, while more difficult, allows the zero-velocity plane to be moved away from the sample boundaries \cite{Derks2004, Derks2009}.  This, in turn, allows particle tracking in bulk, even under very rapid shear conditions.  An alternate approach for torsional flows requires mounting a commercial rheometer on a confocal microscope \cite{Ballesta2008, Besseling2009, Besseling2010, Schmoller2010}.  To achieve a uniform shear rate, commercial rheometers use a cone and plate geometry in which the sample thickness increases linearly with distance from the rotation axis.  Measurements of confined suspensions, however, require a parallel plate geometry in which the sample thickness is uniform but the shear rate is not.  Thus, while this approach allows a great variety of bulk measurements to be paired with simultaneous visualization of sheared microstructures, confined systems under uniform shear cannot be explored using this apparatus.  

The second group of experimental designs use flat, counter-translating surfaces to drive planar Couette flows with straight, parallel streamlines \cite{Haw1998a, Haw1998b, Petekidis2002, Cohen2004, Cohen2006, Solomon2006, Besseling2007, Smith2007, Wu2007, Wu2009}.  
Again, the simplest designs have one surface fixed in the laboratory frame, though independently moving both is possible and has the same advantages as in the analogous rotational design \cite{Wu2007, Wu2009}.  In these planar designs, with careful alignment of the boundaries, sample thicknesses of just a few particle diameters become possible, allowing exploration of thin film flow behaviors associated with confinement \cite{Cohen2004, granick1991motions, israelachvili2011intermolecular}.  However, force measurement is a significant challenge in this flow geometry.  To achieve a high degree of parallelism, shear cell designs tend to use much smaller surfaces than those found in commercial rheometers.  With weak stresses acting on these areas, the resulting shear forces can be on the order of micro-Newtons even when working with fairly viscous suspensions.  For this reason, these planar, uniaxial designs have traditionally been used to study questions concerning particle configuration and dynamics rather than rheological phenomena.

Here, we discuss a new translating, parallel-plate design incorporating both fast confocal microscopy and force measurement \cite{Cheng2011}.  This instrument offers precision control of uniform shear flow.  The plates can be brought within two microns of each other and, with lateral displacements of hundreds of microns possible, very large strains and strain rates can be achieved in strongly confined samples.  Our custom-built, high precision force measurement devices (FMDs) allow \emph{in situ} rheological measurements to be combined with simultaneous visualization of sheared microstructures in this flow geometry.  These devices are sensitive enough to detect shear stresses on the order of tenths of a Pascal.  Moreover, our instrument is capable of performing biaxial shear experiments in which oscillatory shear flows in two orthogonal directions are superimposed.  By adjusting the relative amplitude and phase of these orthogonal components, uniaxial flows in different directions can be explored, as well as perturbations to uniaxial flows and even elliptical flows.  This multi-axial instrument is therefore capable of probing anisotropic structure and dynamics in a variety of systems and, unlike currently available commercial instruments, allows direct visualization of sheared microstructures.

This paper is organized as follows.  In Section II, we discuss design challenges and our solutions to them, focusing on geometric and kinematic aspects of the design.  In Section III, we discuss the force measurement problem in detail and present designs for both uniaxial and biaxial FMDs.  Section IV discusses the use of confocal microscopy in connection with our shear cell design.  Finally, in Section V, we outline the application of our shear cell to the study of some representative experimental systems.

\section{Parallel Plate Shearing Apparatus }

	\subsection{Control of shear cell alignment and spacing}

\begin{figure}[t]
\centering
\includegraphics[width=3.2in]{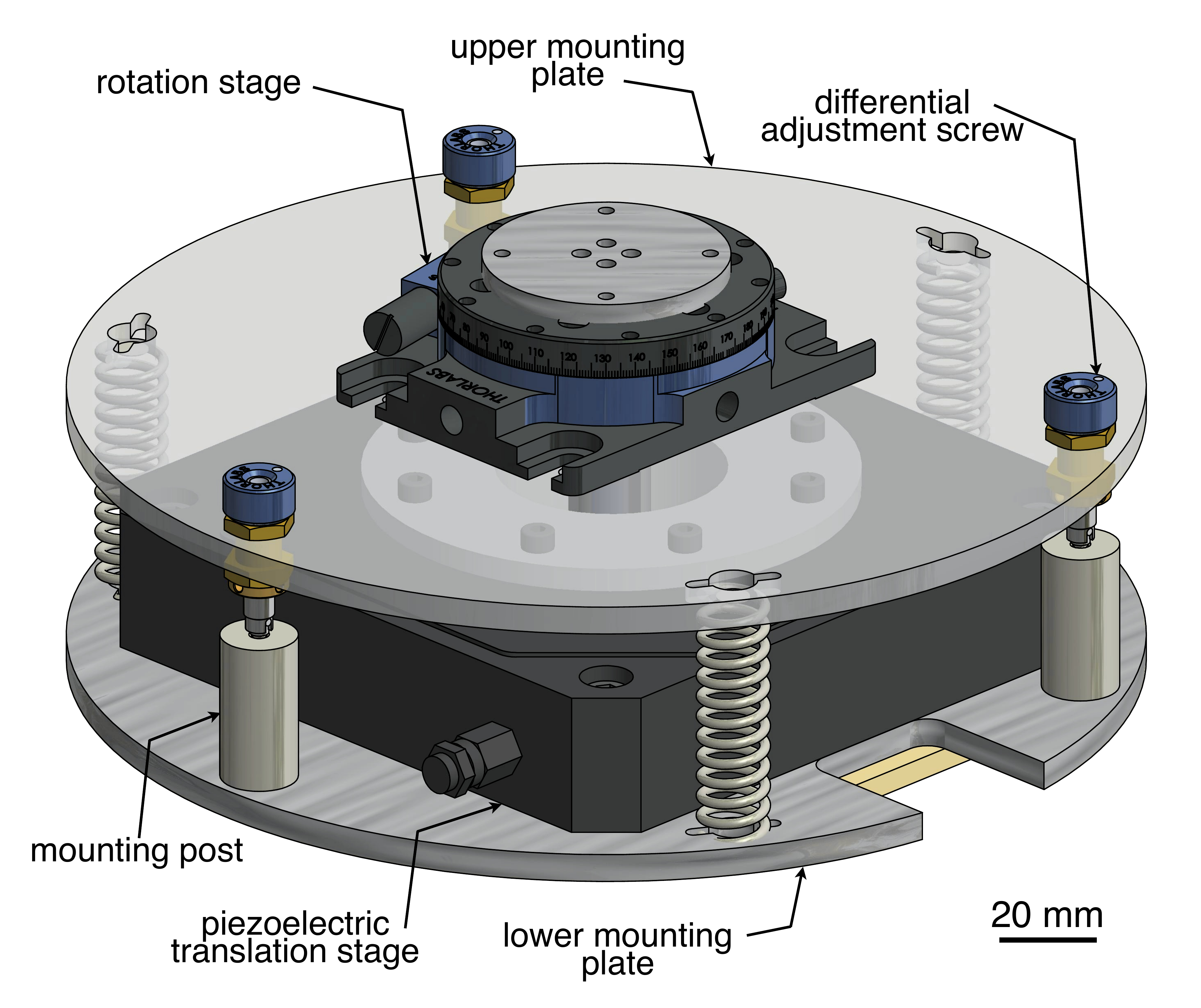}
\caption{Schematic representation of the shear flow apparatus, with semi-transparent rendering of upper mounting plate (angled top view).  The three differential adjustment screws allow us to accurately position the plates in a parallel geometry. The black piezoelectric translation stage drives the bottom plate of the shear cell back and forth, and the top plate of the shear cell and Force Measurement Device are mounted to the top mounting plate. Figures.~\ref{fullapparatus_cut} and \ref{shearcell_cut} show a close-up of the mounting setup and shear cell chamber.}
\label{fullapparatus_angled}
\end{figure}

\begin{figure}[t]
\centering
\includegraphics[width=3.2in]{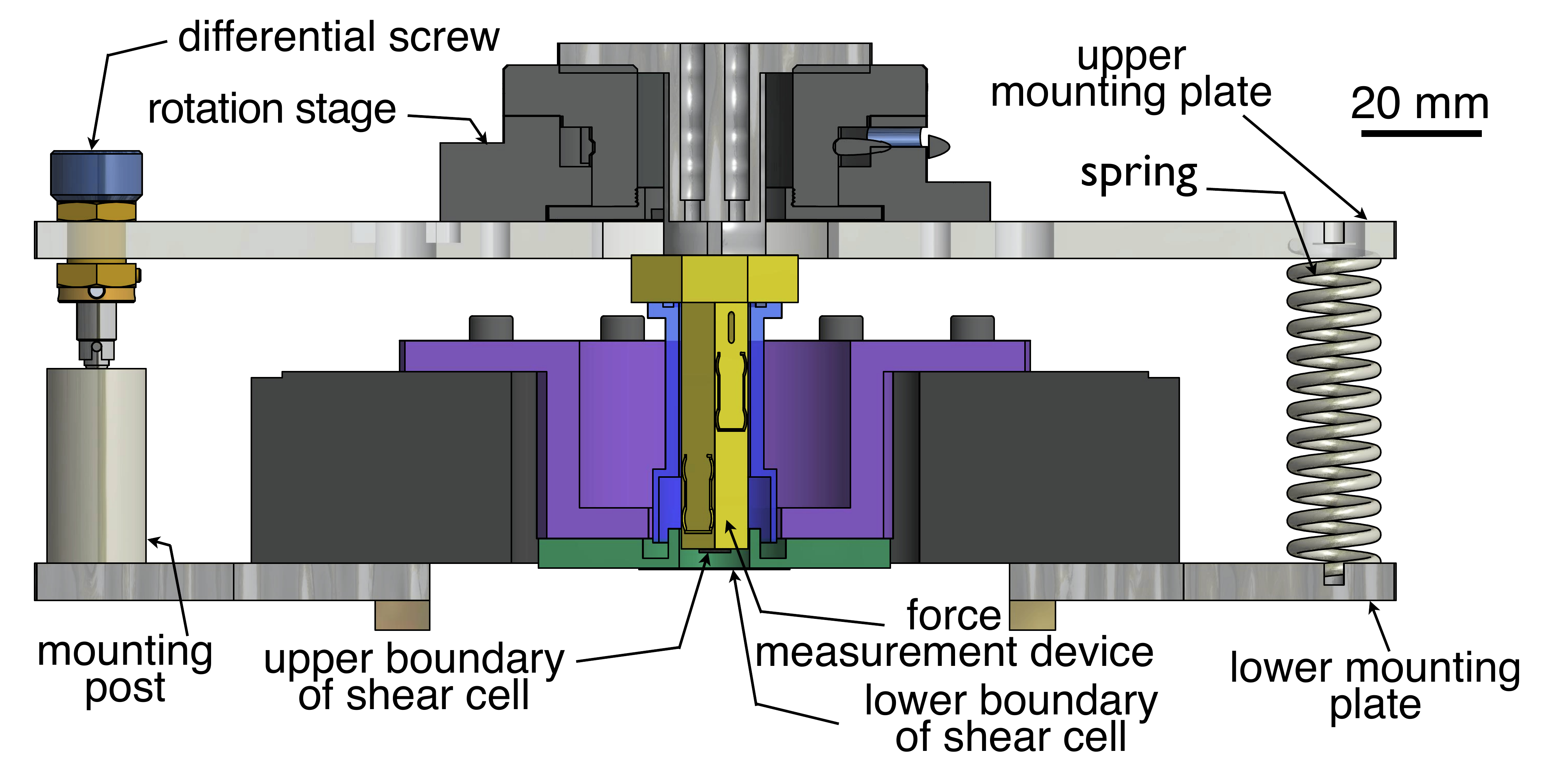}
\caption{Schematic representation of the shear flow apparatus (side view, cut). See Fig.~\ref{shearcell_cut} for an enlarged view of the shear cell.}
\label{fullapparatus_cut}
\end{figure}

The first design challenge is ensuring that the two plates are extremely parallel with a given separation. In our shear flow apparatus, two large mounting plates are used to control the alignment and spacing of the shear cell boundaries.  The lower boundary is attached (via a piezoelectric translation stage) to the lower mounting plate.  The upper boundary is attached via a force measurement device and rotational stage to the upper mounting plate.  This arrangement is shown in Fig.~\ref{fullapparatus_angled} and \ref{fullapparatus_cut}.  To control the spacing and ensure the mounting plates are parallel, we use a constrained system of three adjustment screws, placed at the vertices of an equilateral triangle. Each screw passes through a tight, threaded bushing in the upper mounting plate and has a ball bearing at its tip, which rests on a post attached to the lower mounting plate.  One of these posts has a conical hole in its upper surface, which prevents any lateral motion of its screw.  Another post has a linear groove, which allows lateral motion of its screw but only in one direction.  The last post has a flat surface, allowing its screw to move freely in two directions. Holding the plates together with stiff springs prevents the plates from moving about during shear, keeping the plate separation fixed. The plates are sufficiently thick that the springs cause no observable warping when in place.  

This kinematic mount design exploits the rigid body degrees of freedom of the upper mounting plate. In free space, the plate has exactly three translational degrees of freedom and three rotational degrees of freedom. All three translational degrees of freedom are lost when one of the screw tips engages the conical hole. Two of the rotational degrees of freedom are lost if another screw tip engages the linear groove. The last rotational degree of freedom is lost when the remaining screw tip engages the flat post. Thus, for each setting of the adjustment screws, there is a unique configuration of screw tip positions that immobilizes the upper mounting plate. As the screws are independently rotated, small cooperative motions of the screw tips along the grooved and flat posts allow the mount to freely explore different orientations without building up mechanical stress in either the screws or the plates. In this way, microscopic misalignments due to the mechanics of the mount are minimized.

The adjustment screws and springs are evenly spaced around a circle 200 mm in diameter, visible in the angled view of Fig.~\ref{fullapparatus_angled}. Each screw (ThorLabs, DAS110) has an outer thread for course adjustments and a differential inner mechanism for fine adjustments. The coarse thread advances the screw 0.3175 mm per rotation. Thus, one full rotation of any one of the adjustment screws corresponds to an angular change of only $2.12 \times 10^{-3}$ radians in the relative orientation of mounting plates. The more precise differential adjustment mechanism advances the screw 0.025 mm per rotation. Thus, one full rotation of this mechanism in any one screw corresponds to an angular change of only $1.67 \times 10^{-4}$ radians in the relative orientation of the mounting plates. Since the upper and lower shear cell boundaries are independently fixed to the upper and lower mounting plates, as described above, these adjustments allow precise control of the shear cell geometry.  We find that small cooperative motions using the coarse threads alone are usually enough to align the cell boundaries parallel to within roughly $5 \times 10^{-5}$ radians or, equivalently, to within roughly 0.2 $\mu$m across the entire shear zone.  

The use of three precision adjustment screws not only allows us to level the plates, but it also allows us to adjust their gap over a wide range. With careful alignment of the three screws, the cell boundaries can be brought very close together ({\it e.g.}, 2 $\mu$m), enabling study of samples containing only a few particle layers.  Moreover, due to their large adjustable range, the screws can be adjusted to increase the plate separations to more than 1 mm. 

	\subsection{The shear cell}

\begin{figure}[t]
\centering
\includegraphics[width=3.2in]{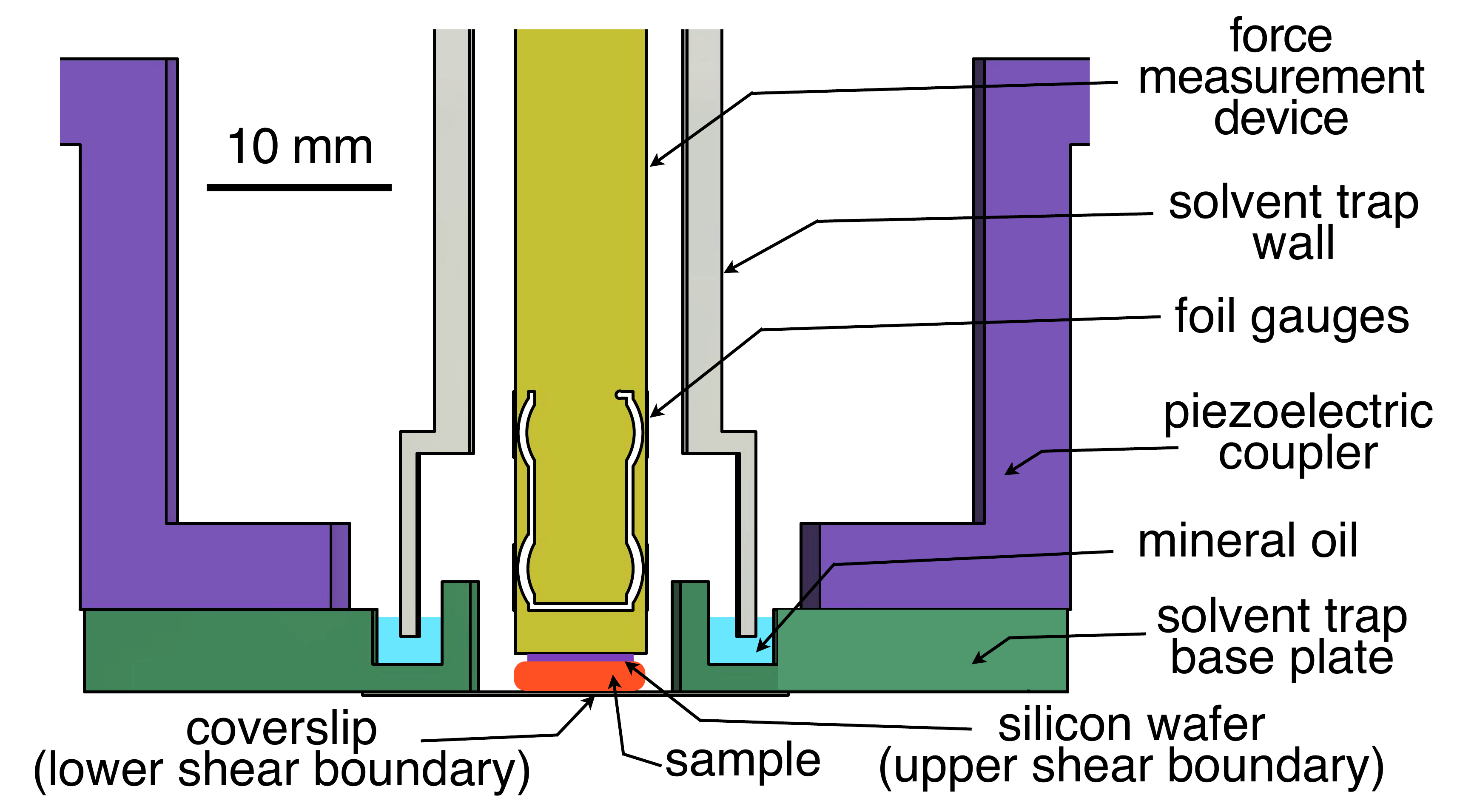}
\caption{Schematic representation showing major components of the shear cell in greater detail (side view, cut). Note that the aspect ratio of the sample pool is exaggerated here to make it visible from the side. In practice, the gap is much smaller relative to the width, as discussed in later sections.  The sample volume can also extend beyond the shear zone, forming a reservoir that is not shown here.
}
\label{shearcell_cut}
\end{figure}

To prevent evaporation and contamination of the sample, the sample loading region is isolated via a solvent trap, as shown in Fig.~\ref{shearcell_cut}. The lower mounting plate is attached to a solvent trap plate with an annular groove; this groove is filled with 300 $\mu$L of mineral oil before beginning an experiment. The top plate has a metal or polycarbonate tube attached to it, via the base of the FMD. This tube provides the lateral sidewalls of the solvent trap. As the top plate is brought down into the sample from above, the bottom rim of this tube lines up with an annular groove in the solvent trap plate.  The 300 $\mu$L pool of mineral oil forms an airtight seal between these two pieces of the solvent trap.  In this way, sample evaporation and exposure to air can be minimized. Since there is no direct mechanical contact between the FMD and the moving pieces of the shear cell, the solvent trap disturbs neither the shear experiment nor the force measurements. 

\begin{figure*}
\centering
\includegraphics[width=5in]{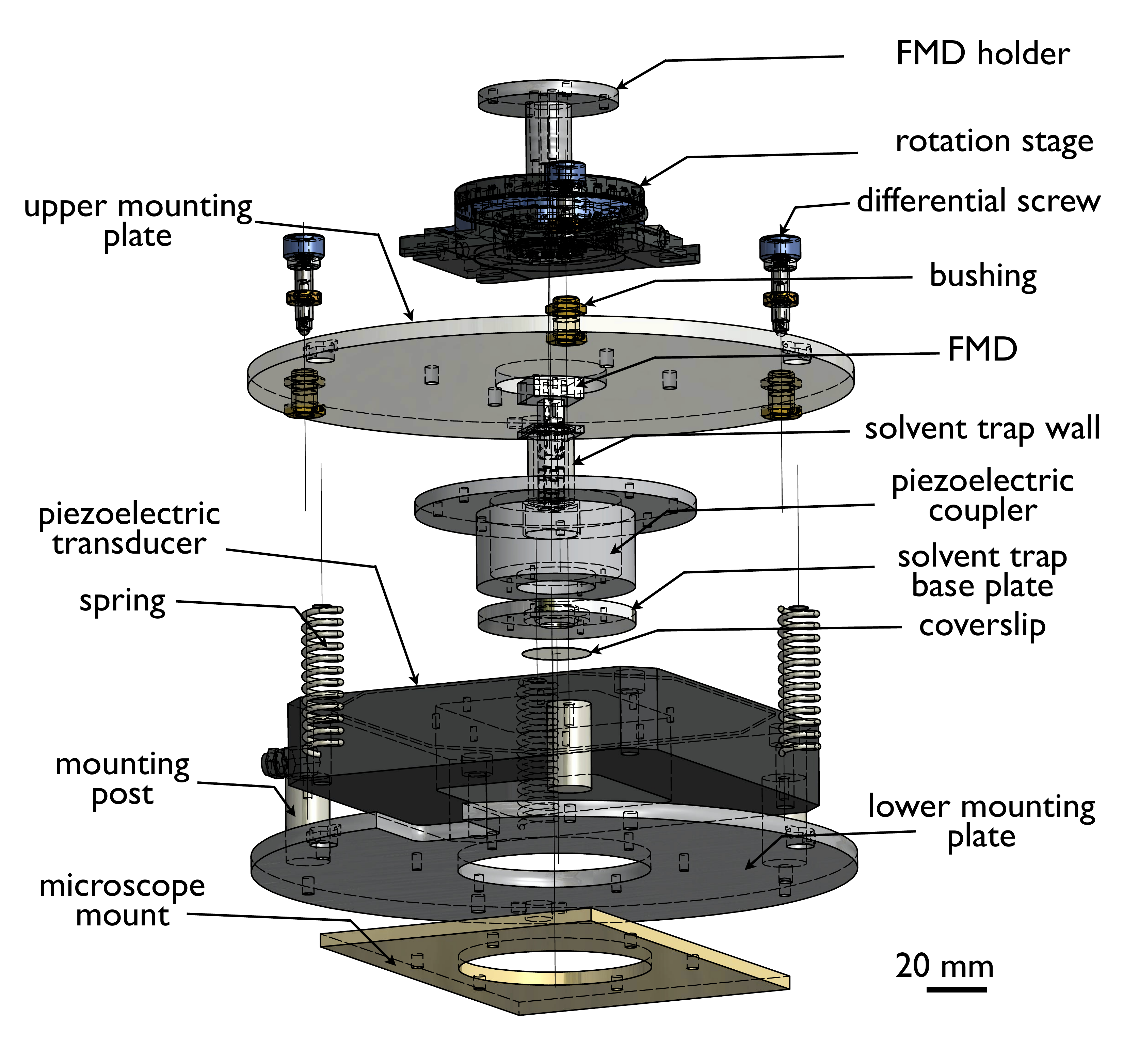}
\caption{Exploded view drawing of the confocal rheoscope.}
\label{explosive_drawing}
\end{figure*}

Our shear apparatus needs a transparent bottom plate to allow confocal imaging, and it needs smooth top and bottom plates to avoid a spatially varying plate separation due to surface roughness. We use a standard No.~1.5 microscope cover slip (Warner Instruments CS-25R15, approximately 170 $\mu$m thick) as the lower boundary of the shear cell. In addition to granting us the ability to image with a confocal microscope, a glass cover slip allows us to treat the surface, \textit{e.g.} by base-washing to clean the surface and to make it hydrophilic, or by silanization to make it hydrophobic. The glass cover slips are both smooth and fairly planar, allowing for a uniform shear profile across the sample. The cover slip is glued to the bottom of the solvent trap plate using ultraviolet-cured optical adhesive (Norland NOA89), as shown in Fig.~\ref{shearcell_cut}.  A half-inch diameter circular hole in the center of the solvent trap plate provides access to the cover slip from above, both for sample loading and for placement of the top plate.  Samples are loaded into the center of the cylindrical space defined by the walls of this hole.  This plate is in turn connected to the lower mounting plate, via a rigid adapter connection to the piezoelectric translation stage. For the upper boundary of the shear cell, we use a 4 mm $\times$ 4 mm silicon wafer, which is atomically smooth.  This wafer is attached to the upper mounting plate via the tip of the force measurement device (FMD). The relationship and order of assembly of the confocal rheoscope parts are illustrated in the exploded view drawing, as shown in Fig.~\ref{explosive_drawing}.

\subsection{Control of shear cell motion}

Finally, our apparatus requires a drive that can operate precisely over a large range of strain amplitudes and shear frequencies.  Drives such as linear actuators or stepper motors provide the ability to access large strain amplitudes, but the precision for typical actuators is around 1 $\mu$m and their frequency range is often limited. For the small plate separations and shear rates we are interested in accessing these limitations are problematic. Thus, for the most recent versions of our apparatus, we selected as our drive a three-axis piezoelectric device with an open central aperture (Physik Instrumente, P-563.3CD). This particular device has a square hole in the center, which allows for access from above for sample loading and for flow visualization from below.  Our piezoelectric device is capable of close-loop travel of up to 300 $\mu$m laterally. For oscillatory shear flow, this translates to a maximum displacement amplitude $u_0$ of roughly 150 $\mu$m. For small gaps, such as $d = 5$ $\mu$m, the maximum strain amplitude $\gamma_0 = u_0/d$ can reach 30 and the maximum strain rate amplitude $\dot{\gamma}_0 = \omega \gamma_0$ can reach 2.1$\times 10^4$.  This value of $\dot{\gamma}_0$ is obtained for an oscillation frequency $\omega/(2\pi)$ of 110 Hz, which can be achieved without approaching the resonant frequencies of the device.  Another important motivation for choosing a piezoelectric controller is the precision. Our  piezoelectric controller has a displacement resolution of 2 nm. Moreover, the piezo displacement is almost  perfectly proportional to applied voltage, with a deviation from linearity of $3 \times 10^{-4}$. Because the piezoelectric transducer design is standard, as newer products come on line these capabilities can be further improved.

This piezoelectric strategy offers flexibility as well as high precision.  Applying a triangular waveform to a single axis of the device, for example, results in steady unidirectional shear with periodic reversals of direction. A sinusoidal waveform, on the other hand, results in an oscillatory shear flow. Thus, the device is capable of many of the controlled motions necessary for standard rheological measurements, including amplitude sweeps, frequency sweeps, and step strains or strain rates, as well as steady shear at constant strains or strain rates.  Moreover, the three-axis capability of our device allows us to select the direction of shear and to shear in multiple directions. 

The superposition of two time-dependent shear flows can be written as:
\begin{equation}
\bm{\gamma}_\textrm{tot}(t) = \bm{\gamma}_1(t) + \bm{\gamma}_2(t) \quad ,
\label{eq:2D}
\end{equation}
where the subscript 1 or 2 indicates the primary or secondary flow and $\bm{\gamma}$ is the tensorial time-dependent shear strain. By imposing different types of shear, we can perform a mixture of creep (square waveform), oscillatory (sine waveform), and continuous (triangular waveform) measurements. If we consider as a special case, a superposition of two oscillatory shear flows, we can write  $\bm{\gamma}_1(t)=\tilde{\bm{\gamma}_1} \sin (\omega_1 t)$ and $\tilde{\bm{\gamma}_2} \sin(\omega_2 t + \delta)$. Here $\tilde{\bm{\gamma}_1}$ and $\tilde{\bm{\gamma}_2}$ are the tensorial strain amplitudes of primary and secondary flows respectively, $\omega$ is the shear frequency, and $\delta$ is a phase difference.  If $\delta = 0$ and $\omega_1 = \omega_2$, then the flow is a uniaxial shear with strain amplitude $\bm{\gamma}_\textrm{tot} = \tilde{\bm{\gamma}_1} + \tilde{\bm{\gamma}_2}$. However if either $\delta \ne 0$ or $\omega_1 \ne \omega_2$, then the applied shear is not uniaxial. For instance, if $\omega_1 = \omega_2$ but $\delta \ne 0$, the shear flow is elliptically polarized. Alternatively, by using a strong primary flow $\bm{\gamma}_1$ to drive the system away from equilibrium and a weak secondary flow $\bm{\gamma}_2$ to probe the system, a biaxial superposition spectroscopy measurement can be performed.
These possibilities open up new avenues of inquiry for investigating the flow behavior of anisotropic materials, including liquid crystals \cite{walker2000orthogonal, volkov1990anisotropic}, colloidal suspensions \cite{osuji2008highly}, polymers \cite{volkov1990anisotropic}, and biolgical tissues \cite{gennisson2010viscoelastic, guo2000spatial} in which the viscoelastic modulus generally varies with the direction of shear.

It is crucial to ensure that the piezoelectric device is aligned with the top and bottom shear cell plates. As discussed above, the relative alignment and spacing of the shear cell's top and bottom plates are determined through adjustments made to the kinematic mount.  In principle, once the cell geometry is set, the motion of the lower boundary of the cell can be computer controlled through the piezoelectric device without any further manipulation of the mount.  This is only possible, however, if the axis of travel initiated by the piezoelectric device has no component perpendicular to the cell boundaries.  Any misalignment of this axis will cause the distance $d$ between the cell boundaries to vary as the lower boundary is moved.  We find that, with sufficiently careful machining of our apparatus components, this problem can be avoided and, in practice, $d$ remains constant to within 0.1 $\mu$m across the device's full range of motion. This small change in $d$ is barely measurable with a confocal microscope, due to intrinsic resolution limits set by the optics.


\section{Force Measurement}

The main challenge in designing a force measurement device (FMD) is obtaining high sensitivities without sacrificing reliability and repeatability and without disturbing the applied shear flow. For instance, to measure the viscosity of glycerol at a strain amplitude of 1 and strain frequency of 1 Hz the FMD needs to be sensitive to stresses on the order of 6 Pa. However, for a 4 mm $\times$ 4 mm wafer, this corresponds to forces on the order of 0.1 mN. To measure the viscosity of water requires the FMD to resolve forces as small as 0.1 $\mu$N.  Thus, {\it in situ} measurements using our shear cell require extremely sensitive devices.

	\subsection{Uniaxial force measurement devices}
	
\begin{figure}
\centering
\includegraphics[width=3in]{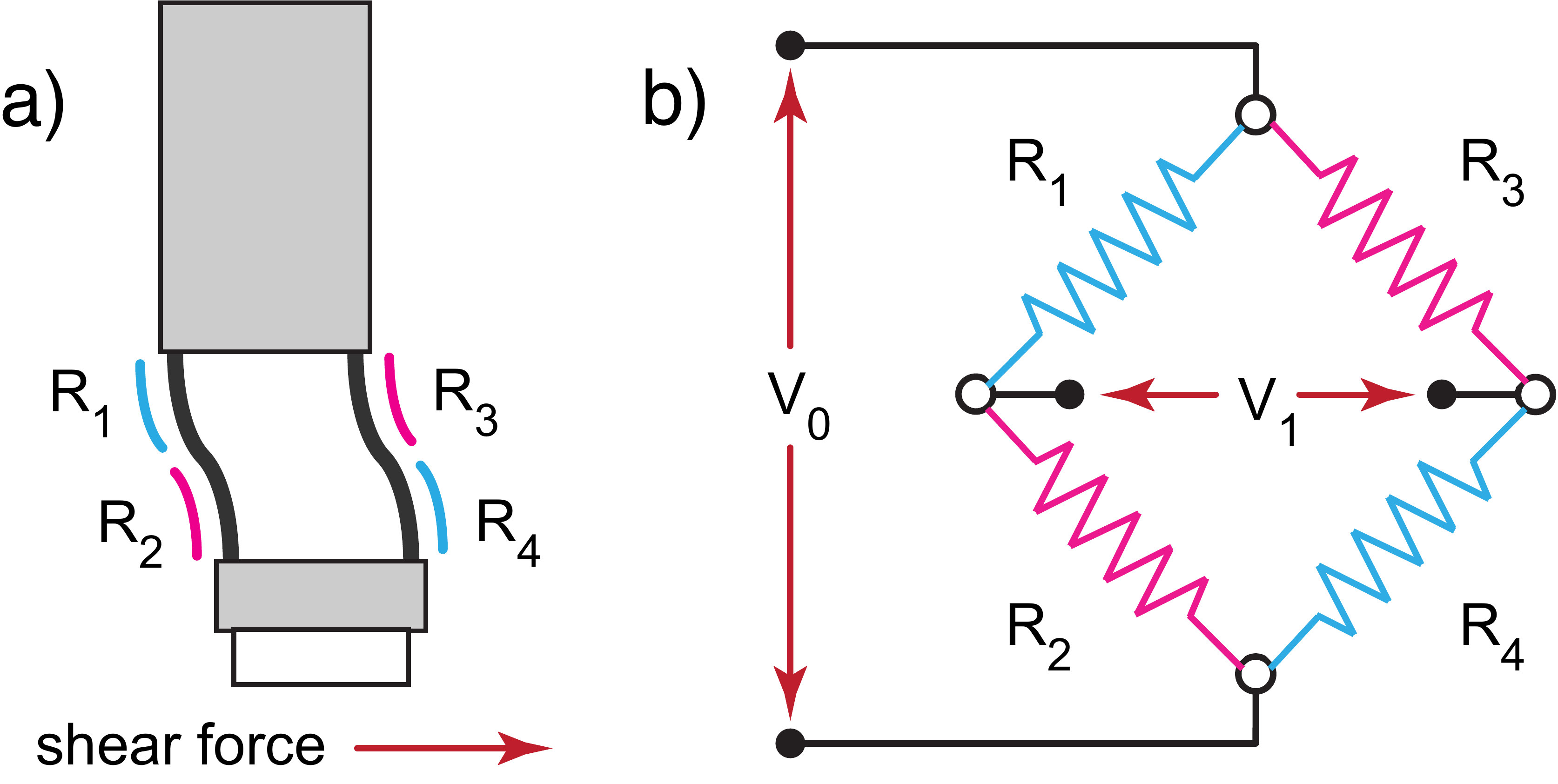}
\caption{(a) Schematic representation of force measurement device under shear (side view), showing exaggerated deformation of the cantilevers and placement of the strain gauges.  Note that for this deformation pattern $R_1$ and $R_4$ experience inward curvature, while $R_2$ and $R_3$ experience outward curvature. (b) Wheatstone bridge circuit configuration.}
\label{wheatstone}
\end{figure}

Our FMDs operate by converting laterally-oriented shear stresses into small but measurable deflections in a system of cantilevers.  Our uniaxial device consists of a thin, horizontal plate suspended from a pair of parallel cantilevers oriented vertically, as shown in Fig.~\ref{shearcell_cut}.  The entire device is machined from a single block of aluminum.  The silicon wafer forming the upper boundary of the shear cell is anchored to the tip of the device.  Laterally-oriented shear stresses deflect the wafer horizontally, creating a characteristic sigmoidal distortion in the cantilevers, which is shown schematically in Fig.~\ref{wheatstone}a.  The vertical deflection associated with these distortions is negligible, ensuring that the alignment and spacing of the shear cell boundaries are preserved under shear. 

The deflection of the cantilevers tells us the force exerted on the top plate. We measure this deflection in the double cantilevers using a system of strain gauge resistors, which are typically foil or semiconductor gauges. The resistance of these strain gauges increases when the strain gauge is stretched; the change in resistance is proportional to the strain in the resistor. Two of these strain gauges are placed on the outer surface of each cantilever, as shown in Fig.~\ref{wheatstone}a.  When a cantilever is deflected, the inward curvature of one of the strain gauges causes its resistance to increase, while the outward curvature of the other strain gauge causes its resistance to decrease.  

Wiring all four strain gauges in a standard Wheatstone bridge configuration provides a sensitive method for measuring these changes \cite{branan2002rules}.  This circuit, shown in Fig.~\ref{wheatstone}b, requires a excitation voltage $V_0$ across the bridge in one direction.  For small changes in resistance and four strain gauges of equal resistance $R$ in equilibrium, the voltage $V_1$ measured across the bridge in the other direction has the approximate form
		\[  V_1 \approx \frac{V_0}{4} \big( \frac{ \Delta R_1}{R} - \frac{\Delta R_2}{R} 
			- \frac{\Delta R_3}{R} + \frac{\Delta R_4}{R} \big) \quad .  \]
The circuit is wired so that all four gauges make complementary contributions to the measured voltage $V_1$.  In particular, for the distortion pattern shown schematically in Fig.~\ref{wheatstone}a, the two strain gauges colored cyan have the same effect:  both experience a positive $\Delta R$ and, by assigning these to positions 1 and 4 in the bridge circuit, both make a positive contribution to $V_1$.  Likewise, the two strain gauges colored magenta both experience a negative $\Delta R$ and, by assigning these to positions 2 and 3 in the bridge circuit, both make positive contributions to $V_1$ as well.
Assuming a symmetric distortion pattern, these contributions are all equal in magnitude, $| \Delta R_j | = \Delta R$, and the above equation reduces to
		\begin{equation}  \label{Wheatstone}
			V_1 \approx V_0 \frac{ \Delta R}{R} \quad .
		\end{equation}
Thus by measuring the voltage $V_1$ across the bridge, we can determine the deflection of the strain gauges and thus the deflection in the FMD.  We amplify $V_1$ using a Vishay 2310 signal conditioning amplifier with a gain of $10^4$.  To avoid distortions in the output signal we refrain from any filtering of the signal.  This conversion of applied stresses into measurable signals is summarized schematically in Fig. 7, noting the experimental component or process responsible for each step. 

\begin{figure*}
\centering
\includegraphics[width=6in]{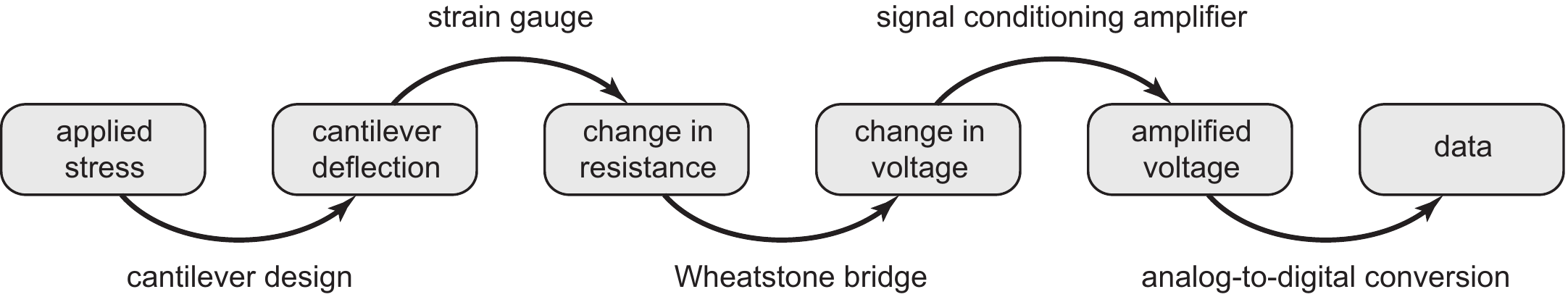}
\caption{
Data flow in our force measurement device.  Shear flow in the sample exerts a stress on the end of the FMD, resulting in a lateral deflection of the cantilevers.  Curvature associated with this deflection changes the resistance of four strain gauges mounted on the cantilevers.  Using a Wheatstone bridge circuit, these resistance changes are converted to a change in voltage.  However, due to the small magnitude of the strains involved, we must amplify the corresponding voltage change by using a signal conditioning amplifier. Finally, the amplified voltage is output to a computer using an analog-to-digital converter. 
}
\label{sensitivity}
\end{figure*}

	\subsection{Uniaxial calibration and sensitivity measurements}

As discussed above, we expect that the measured voltage $V_1$  should vary linearly with the deflection of the FMD. For small deflections, the FMD deflection is proportional to the applied stress. Thus the output voltage $V_1$ should be proportional to the force on the FMD as well. To characterize the response and the performance of our force measurements, we conducted two types of calibration:  first, by applying a known force to the FMD and measuring the output signal, and second, by mounting the FMD on the shear cell and shearing a sample of known viscosity. 

\begin{figure}[!t]
\centering
\includegraphics[width=3.5in]{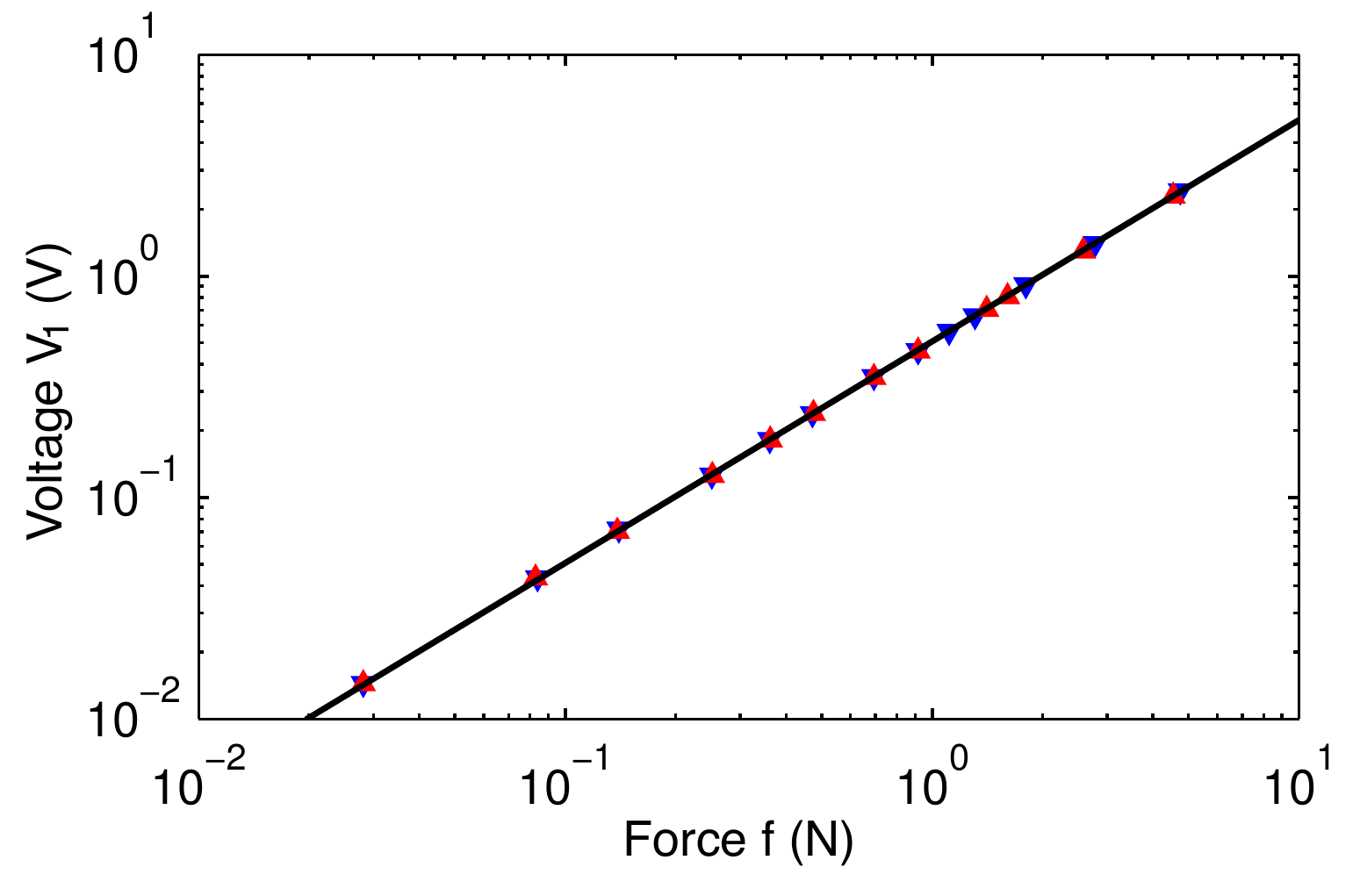}
\caption{Calibration data establishing that voltages obtained from a uniaxial force measurement device are proportional to lateral forces applied to the end of the device.  Data obtained for two opposite orientations (blue and red triangles) fall onto the same line:  the device is both linear and symmetric.  The slope of this line is $0.506 \pm 0.002$ V/N.}
\label{fmdcalib}
\end{figure}

We first calibrated the FMD by applying a known force to it and measuring the output signal. We mounted the FMD sideways, so its measurement axis is aligned vertically instead of horizontally. Then, we hung small weights from the end of the FMD and measured the voltage $V_1$ across the Wheatstone bridge. The raw voltages from the device were amplified using a signal conditioning amplifier and logged digitally using LabVIEW.  Flipping the device over and repeating this procedure, we obtain the voltage response $V_1$ for forces applied in the opposite direction. The data in both directions, with a range of different weights varying from $f=10^{-2}$ to $10^1$ N, are plotted in Fig.~\ref{fmdcalib}.  A linear fit to the data, with the slope as the only free parameter, produces the line shown in the Figure. Note that data obtained from both orientations of the device fall onto the same line.  The slope of this line, $V_1/f=0.506 \pm 0.002$ V/N, provides a calibration factor with which we can convert measured voltages into forces. By dividing the forces by the known area of the silicon wafer, we can obtain the average stress on the FMD in the direction of the measurement axis, defined by the cantilevers.  Thus our device provides quantitative access to stresses in the sample.  The rotation stage on the upper mounting plate allows us to rotate the FMD and to align its measurement axis with any desired in-plane direction. 

For steady uniaxial shear, the sensitivity of the device is limited by noise in the measurement system.  With excitation voltages $V_0$ on the order of 5V, our signal conditioning amplifier introduces noise levels on the order of a few mV.  This makes shear forces of less than roughly 10 mN difficult to measure precisely.  For oscillatory shear, however, we can use Fourier analysis to detect much smaller force signals buried in the noise.  The Fourier transform of the force signal shows a definite peak at the oscillation frequency, which can be converted to a force amplitude.  This approach, which is the motivating principle behind the lock-in amplifier, is limited by the noise floor in the signal's Fourier transform.  This typically gives us access to force signal amplitudes as small as 20 $\mu$N, \emph{i.e.}, three orders of magnitude beyond the limit associated with the noise floor in real space. Thus, oscillatory shear is especially useful for experiments in which we expect very small forces.

\begin{figure}[!t]
\centering
\includegraphics[width=3.2in]{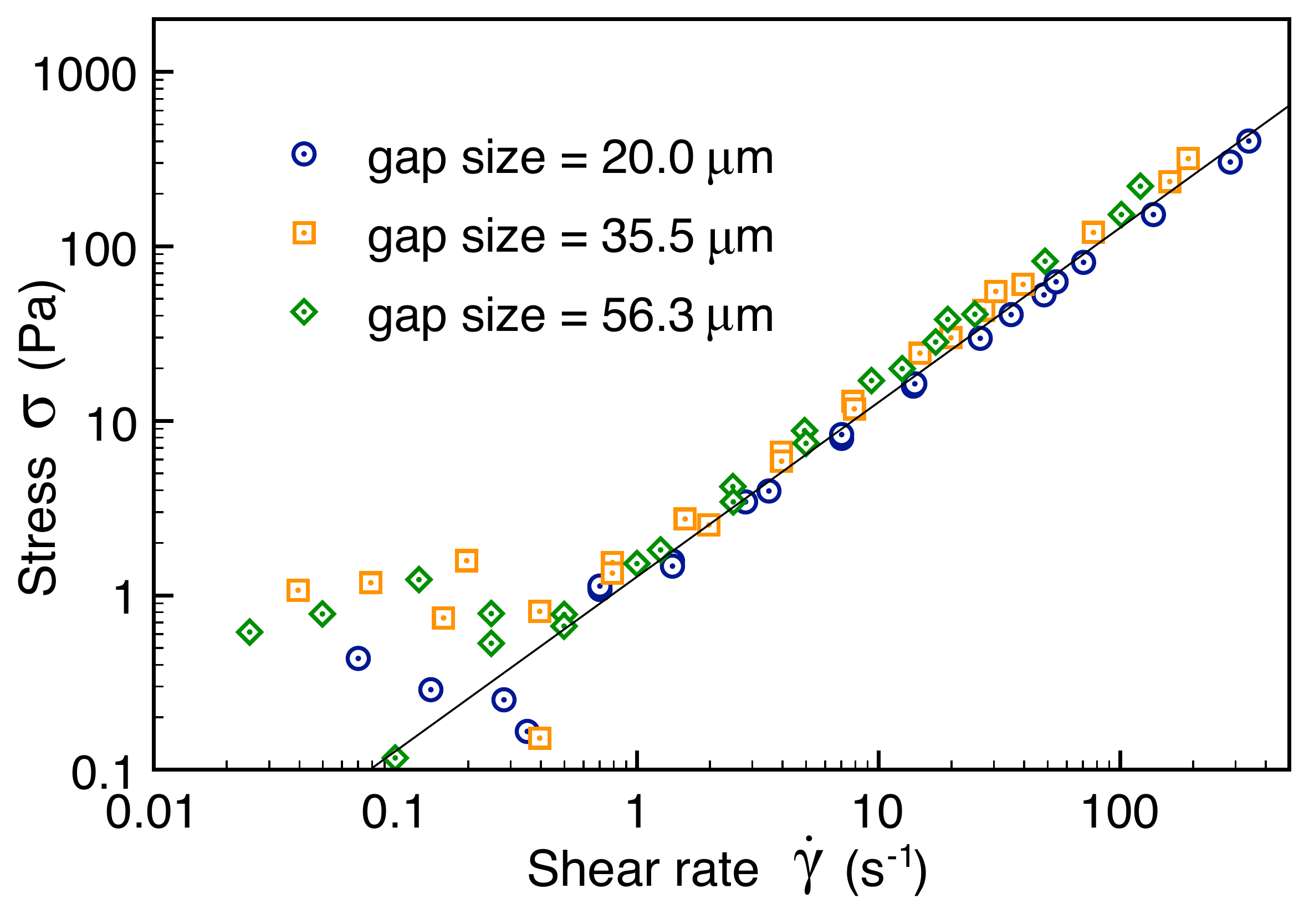}
\caption{Flow curve for glycerol obtained using our apparatus.  Six different data sets, taken at a range of gap heights and strain amplitudes, collapse onto a common line of slope $\eta = 1.28 \pm 0.07$ Pa$\cdot$s.  This value falls within the range of accepted values for the viscosity of glycerol.}
\label{glycerol} 
\end{figure}

To test the FMD using a sample with known viscosity, we characterized the flow curve of glycerol. The results are shown in Fig.~\ref{glycerol}.  A small quantity of fluorescein is mixed into the glycerol, enabling us to use the confocal microscope to measure the gap heights.  For each of three different heights, we sweep through a range of frequencies between 0.05 Hz and 100 Hz for two different strain amplitudes $\gamma_0$ and measure the force amplitude using Fourier analysis.  The sample is allowed to equilibrate with the air environment inside the solvent trap for 30 minutes before beginning shear.  Plotting stress amplitude versus strain rate amplitude, the data collapses onto the straight line shown in Fig.~\ref{glycerol}.  The slope of this line obtained from a least squares fit to the data, $\eta = 1.28 \pm 0.07$ Pa$\cdot$s, agrees with accepted viscosity values obtained under similar temperature and humidity conditions.  The scatter at the base of the curve reflects a combination of noise in the spectrum and finite resolution of our data acquisition board.  

Careful consideration of the conversion process summarized in Fig. 6 helps us better understand which aspects of our design most severely limit the sensitivity of the device.  Cantilever design is clearly an important factor, for example.  Using finite element calculations and Euler-Bernoulli beam theory, we confirm that under applied shear stresses both of the cantilevers in our device assume the characteristic curved shape shown schematically in Fig.~\ref{wheatstone}a.  For the weakest stresses resolved in in Fig.~\ref{glycerol}, however, the tip deflection turns out to be extremely small, {\it i.e.}, less than the lattice constant of aluminum.  At these stress levels, the resulting resistance changes in the four strain gauges are also quite small and subsequent amplification by the Wheatstone bridge circuit and signal conditioning amplifier produces a signal which is buried in the noise.  As emphasized above, however, Fourier techniques can be used when dealing with oscillatory shear flows and these techniques extend the device's sensitivity considerably.  Even for an oscillating shear stress amplitude of only 3 Pa, there is an unambiguous peak at the oscillation frequency in the measured signal's Fourier spectrum.  Thus the noise floor is not the strongest limitation.  Moreover, additional gain amplifies both the signal peak and the noise, without improving their ratio.  Therefore, additional amplification will not increase our sensitivity either.  Likewise, the finite bit depth of the analog-to-digital conversion process is not a major limitation when smaller signals are expected and the range of the data acquisition card is adjusted accordingly.  Apart from the noise floor in the Fourier spectrum, cantilever and strain gauge design emerge as the most significant factors limiting sensitivity.

In summary, our current uniaxial device is capable of resolving surprisingly small deflections and is sufficient for rheological studies like those described in the Applications section.  Access to even smaller shear stresses, however, requires thinner cantilevers and more sensitive strain gauges.  These insights motivated the design of our biaxial force measurement devices, described in the next section.

	\subsection{Biaxial force measurement devices}	

To construct a biaxial Force Measurement Device, our design combines two independently functioning uniaxial FMDs. However, combining the two FMDs into one device introduces two new major design challenges. The first challenge is to combine the FMDs in a geometry that can  measure the same region of the sample, while still fitting into the sample testing chamber. Moreover, the implementation of this design geometry must not decrease the sensitivity of either axis of the FMD. The second major design challenge is to eliminate the coupling between the signals from the two different axes. For instance, while the sample is being sheared along one axis, the other axis of the FMD should have zero signal output. 

\begin{figure}[!t]
\centering
\includegraphics[width=3.4in]{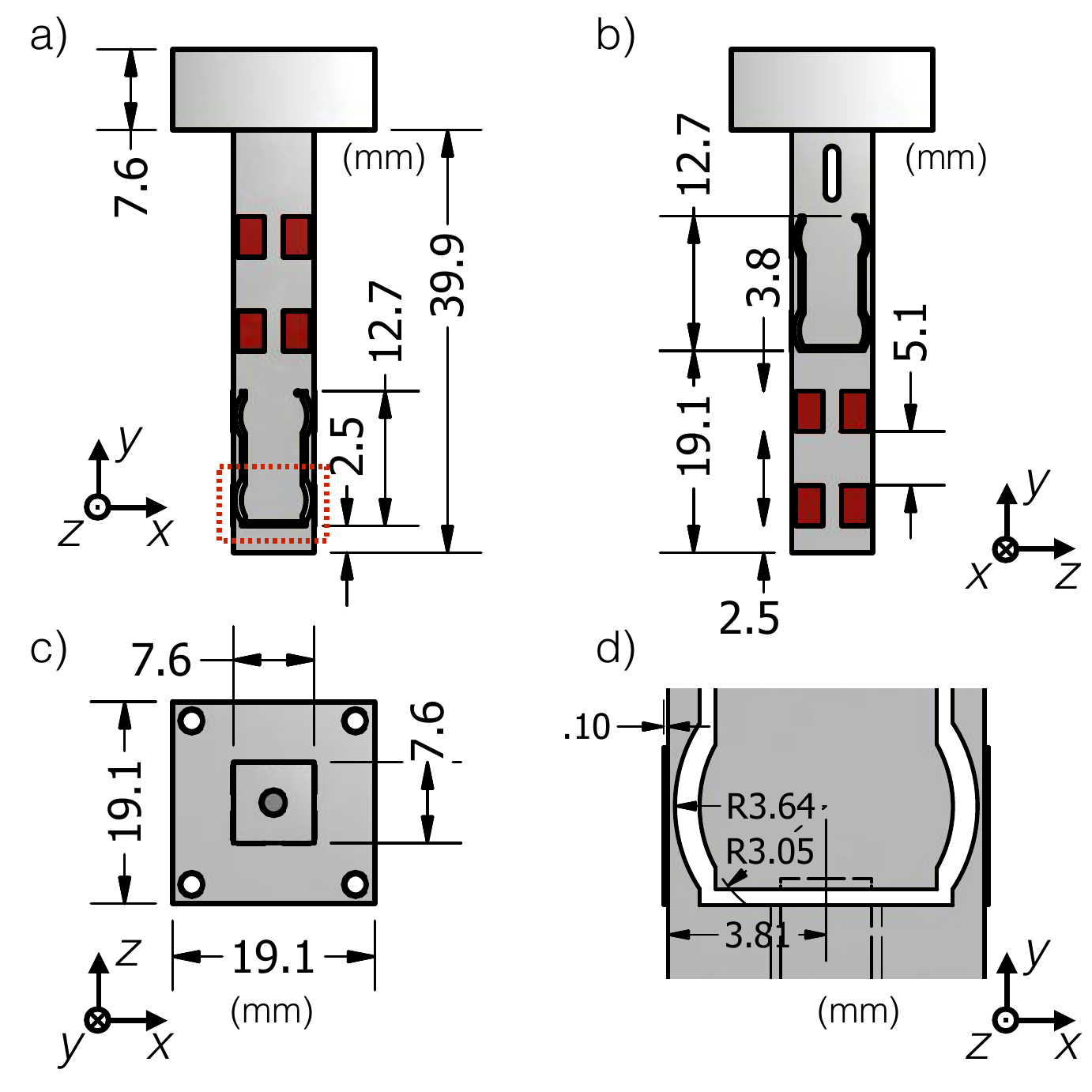}
\caption{Schematics representation showing major dimensions of the biaxial force measurement device from (a) side (b) back and (c) top views. The dimensions of the cantilevers and foil gauges (red box in (a)) are labeled in (d).}
\label{FMD_detail_dimension} 
\end{figure}

\begin{figure}
\centering
\includegraphics[width=3.2in]{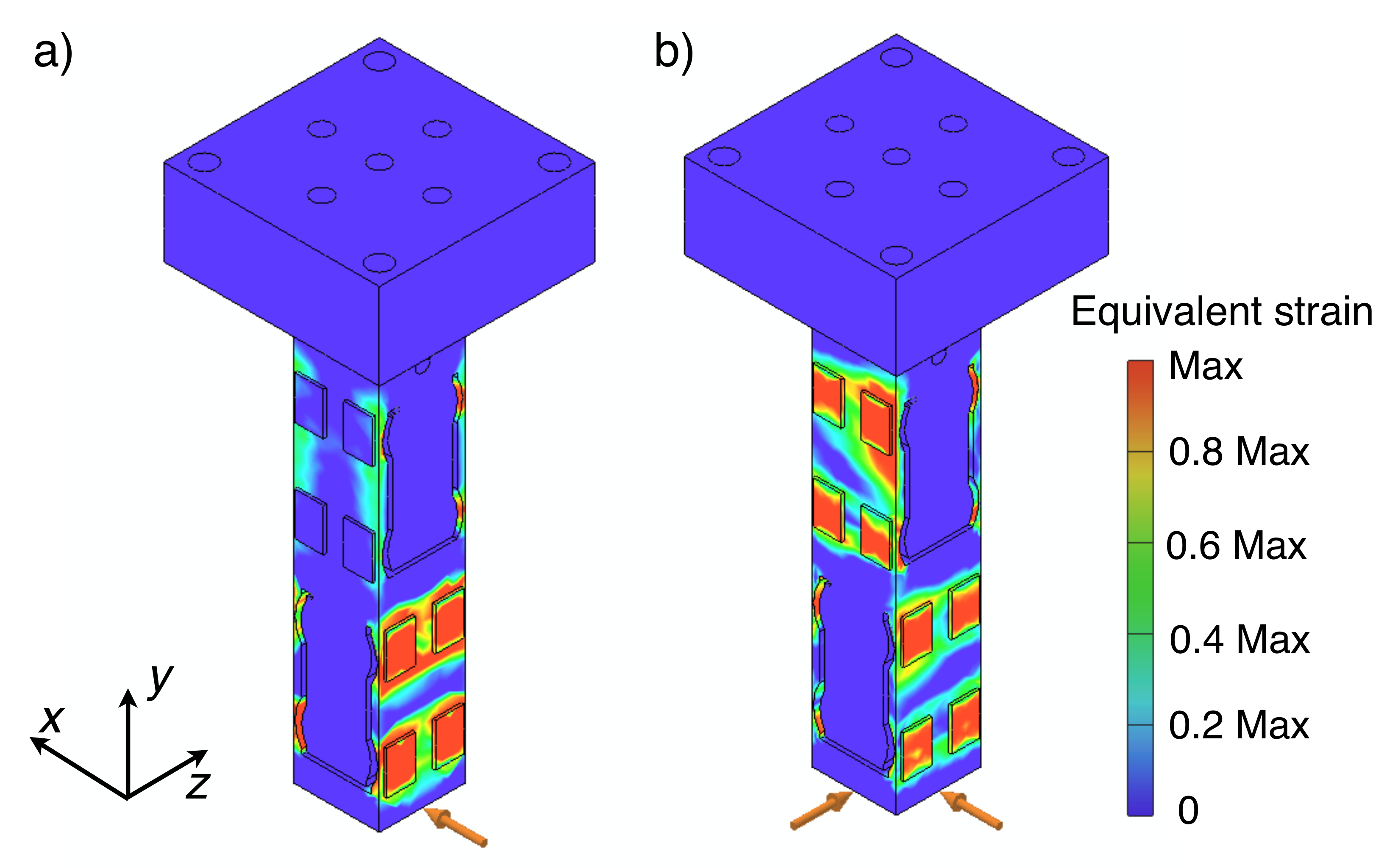}
\caption{Finite element analysis of the biaxial force measurement device with forces (a) perfectly aligned along the bottom cantilever's axis, and (b) at 45$^\circ$ to both cantilevers' axes. The color bar illustrates the value of equivalent strain, which is the magnitude of the stain tensor. As the shear force is applied along the $x$-axis, most of the strain is focused around the strain gauge area of the $x$-channel (a). Likewise, as the force is applied along both $x$-axis and $z$-axis simultaneously, the gauges of both channels sense large strain at the same time. (b)}
\label{biaxial_FEA}
\end{figure}

We thus place the second FMD in series with (\textit{i.e.}, vertically on top of) the first, orienting the uniaxial FMDs at 90$^\circ$ relative to each other as shown in Fig.~\ref{biaxial_FEA}.  By including a solid block between the two, the boundaries of each cantilever remain clamped, as in the uniaxial version of our FMD. When a shear force is applied on the biaxial FMD, the identical boundary conditions ensure that the mechanical response of the double cantilevers is the same as that of the uniaxial device. The vertical arrangement of the uniaxial FMD designs allows the force measurements to be taken at the same location in the sample. Moreover, since the biaxial and uniaxial FMDs' widths do not differ, the new biaxial FMD still fits into the sample testing chamber. Some major numerical quantities on the biaxial FMD physical dimensions are provided in Fig.~\ref{FMD_detail_dimension}.

To convert the deflection of the cantilevers into an electrical signal, we mount eight strain gauges -- four for each of the two axes -- on the biaxial FMD. For ease of mounting, we place all the strain gauges on one side of the cantilever; our symmetric design ensures that this single-sided arrangement does not cause a decrease in the FMD's performance. We then wire these strain gauges into two independent Wheatstone bridges, one for each axis. The wiring for each axis is the same as the wiring for the uniaxial FMD. To minimize the number of wires we let the two Wheatstone bridges share the same excitation voltage $V_0$. The signal voltages, $V_{1x}$ and $V_{1z}$, are then passed via one digital acquisition card to a computer. By using two separate Wheatstone bridges, we can convert the stress response along each axis of the FMD into a separate electrical signal. 

In general, a force on the bottom plate will deflect both the upper and lower sets of cantilevers in our biaxial FMD. The cantilevers are thin rectangular plates, and deflect most easily along the thin axis. However, exerting a force along the \textit{thick} axis -- orthogonal to the thin axis -- will still deflect the cantilever parallel to the force, albeit by a much smaller amount. The ratio of these two deflections will be the quotient of the corresponding bending moments.  
The bending moments along the thin and thick axes are $\propto h^3b$ and $\propto b^3h$, respectively \cite{Landau}.  Thus their quotient is the square of the cantilever aspect ratio, $(h/b)^2$. To minimize the mechanical coupling between the FMD's two axes, we design the cantilever with the smallest possible value of $h/b$. For our design, the ratio of the bending moments $I_x/ I_z = (h/b)^2 = 2.5 \times 10^{-3}$, where $h$= 0.5 mm and $b$= 10 mm. Thus our design has an extremely small coupling. To understand the coupling in more detail, we performed a finite element analysis, displayed in Fig.~\ref{biaxial_FEA}. As expected, we find a small but nonzero mechanical coupling between the two sets of cantilevers. While our design significantly reduces the mechanical coupling between the axes, the coupling is still nonzero and must be accounted for. 

There is a secondary source of coupling in the FMD, due to the electronic data acquisition and signal amplification circuits. If the signal amplifier impedance is high (\textit{e.g.}, resulting from a low-pass electronic filter), the response time of the digital acquisition card can be dramatically increased. Thus, when the digital acquisition card switches from reading out one axis to the other, the residual signal from the previous measurement dominates the new measurement. These residuals cause an additional electronic coupling between the two force measurement axes \cite{taylor1997data}. To avoid this issue, we use two separate signal amplifiers with extremely low impedance and avoid filtering the signal. As with the single axis FMD post-processing of data from an oscillatory force measurements via Fourier transforms can further enhance the measurement sensitivity. 

Finally, an additional source of apparent coupling can arise due to misalignment between the FMD's and the piezo's axes. If the force measurement axes are misaligned with the displacement axes, then a motion which is intended to be along one axis of the FMD will actually have components along both of the FMD's axes. This will result in an apparent force perpendicular to the applied flow.

	\subsection{Biaxial calibration and sensitivity measurements}

We first calibrate the alignment of the FMD's axes, relative to the piezo's axes. To do this, we mount the biaxial FMD on the shear apparatus via a rotation stage. We then apply a uniaxial shear flow and finely adjust the angle of the rotation stage, maximizing the signal in one channel and minimizing the signal in the other. We double-check the alignment by switching the direction of the applied flow by 90$^\circ$. When the force measurement orthogonal to the shear flow is minimized for both flow orientations, then the biaxial FMD is aligned. 

\begin{figure}
\centering
\includegraphics[width=3in]{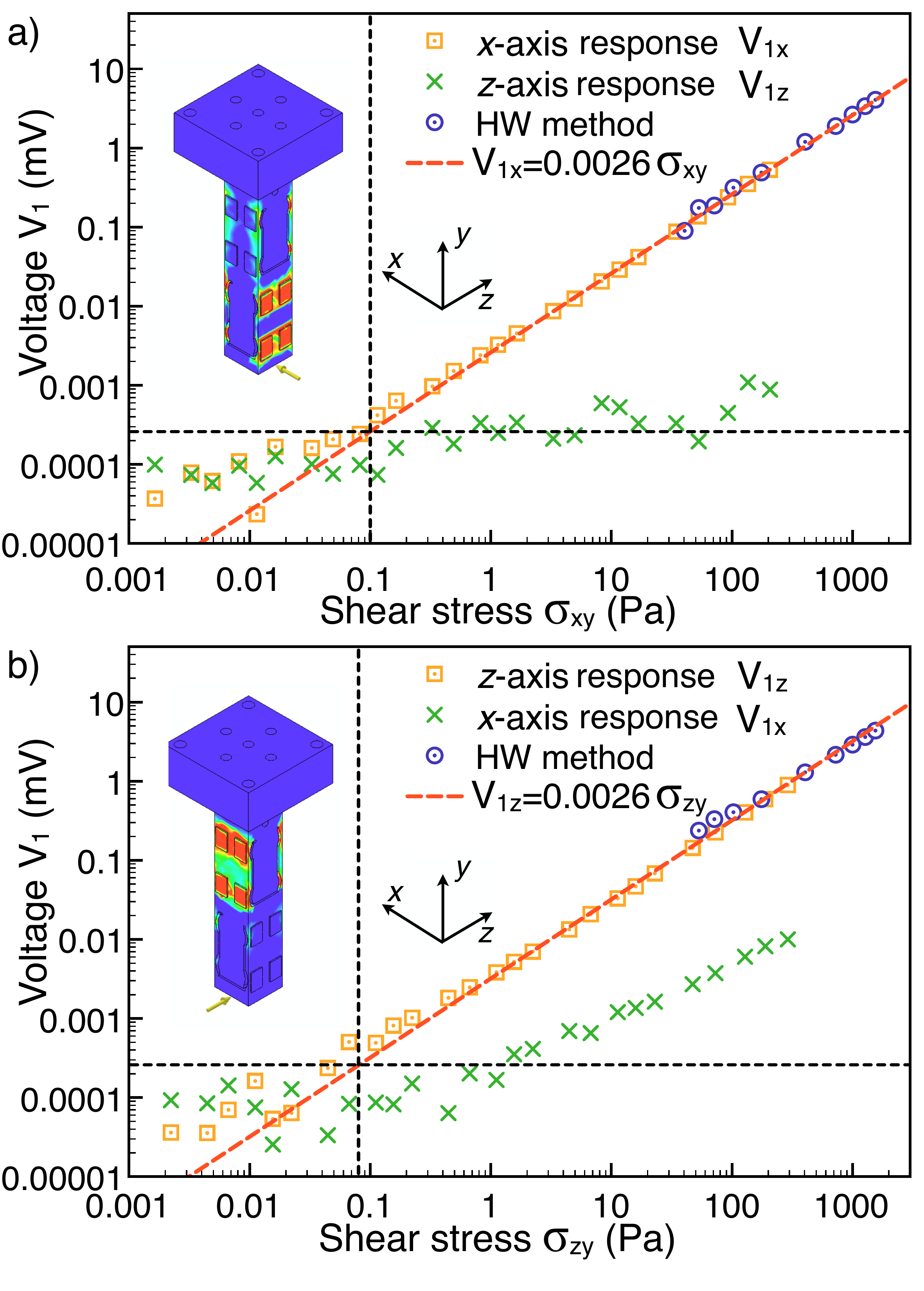}
\caption{The shear stress calibration with shear cell experiment and hanging weight method for $x$-axis (a) and $z$-axis (b). The orange data represent the response of the axis that is along the shear direction, and the green data represent the response of another axis. The blue data are the voltage response measured with the hanging weight method. The red curves are the linear fit to the data. The schematics of the strained force measurement device are generated by the finite element analysis. [N. Lin \textit{et al., Soft Matter}, DOI:10.1039/C3SM52880D]\cite{lin2014} -- Reproduced by permission of The Royal Society of Chemistry}
\label{biaxial_calibration}
\end{figure}

To experimentally determine the coupling between the FMD's axes, we performed force measurements on a sheared standard-viscosity liquid (VIS-RT5K-600, Paragon Scientific). Since the fluid viscosity is known, the stress on the FMD is known for a given shear rate. By varying the strain amplitude at a fixed shear oscillation frequency $f = 10$ Hz, we measured the performance of the FMD over six orders of magnitude in stress, ranging from $1\times 10^{-3}$ to $2\times 10^3$ Pa. Figure~\ref{biaxial_calibration} shows the resulting shear stress versus FMD output voltage, for shear flows oriented along both the $x$-axis (a) and $z$-axis (b) of the biaxial FMD. While there is little signal along the $z$-axis when the flow is along $x$ (Fig.~\ref{biaxial_calibration}a), we do find a measurable force signal along the $x$-direction when the flow is strictly along $z$ (Fig.~\ref{biaxial_calibration}b). The calibration shows that this $xz$-coupling is around $\sim1\%$. Since we reduce the thickness of the cantilevers by factor of three in the design of biaxial FMD, the stress resolution of this new device is approximately ten times higher than the previous uniaxial design.

What is the source of this coupling? Our signal amplification and digital acquisition circuit designs have eliminated the possibility of electronic coupling between the two axes. Figure~\ref{biaxial_calibration}a shows that the system has been aligned as best as possible; thus this coupling cannot result from the misalignment of the FMD with respect to the piezoelectric stage. A clue to the origins of the residual coupling lies in the fact that the $zx$-coupling is zero while the $xz$-coupling is finite. This implies that the coupling matrix is not symmetric, and therefore that its principle axes are not orthogonal. Thus, we speculate that the residual coupling arises from imperfect installation of the strain gauges. Any asymmetry in the strain gauge installation will couple with the cantilever's transverse mechanical deflection. In practice this results in a residual coupling and a non-symmetric coupling matrix. It is a testament to the quality of the strain gauge installation that the coupling is only at the 1\% level. 

Although the coupling between the two axes of the FMD is extremely small, ideally we would like to avoid it altogether. To further eliminate the coupling, we can proceed in two ways. First, we can use the coupling matrix's inverse to convert the measured voltage signal into actual forces. Alternatively, we can apply shear flows in two orthogonal directions with different frequencies. Then, Fourier analysis can be used to pick up the response only at the applied frequencies. This allows for easy disentanglement of the stress responses from the different applied flows. 

Shearing the viscosity standard liquid also provides us with measurements of the FMD's sensitivity. The data in Fig.~\ref{biaxial_calibration} demonstrate that both channels of the biaxial FMD have similar force sensitivity and response. The voltage signals respond linearly with the applied stress over the entire measured range. The biaxial FMD is sensitive to stresses down to $\approx 0.1$ Pa, at which point noise starts to dominate the signal. At force values higher than the range accessible with the viscosity standard liquid, we verify the calibration by hanging weights off the FMD. We find that the viscosity standard calibration and the hanging weight calibration are in excellent agreement over the overlapping range. Thus the FMD is linear and accurate over at least four decades in applied stress. Moreover, our FMD was also tested during the strain gauge installation and was shown to behave linearly with the same constant we measure for forces up to 15 N. This force would correspond to $6\times10^{5}$ Pa. Thus, our FMD functions linearly and accurately over a range of more than six decades.

\section{Visualization and Flow Characterization}

For over a decade, confocal microscopy has been recognized as ideally suited to the quantitative study of soft materials  \cite{Prasad2007, Dinsmore2001, jenkins2008confocal, besseling2009quantitative}. Our central purpose in designing a shear apparatus and force measurement device is to use them in conjunction with a high-speed confocal microscope. This allows us to correlate the real-time evolution of the sample's three-dimensional microscopic structure with its rheology and mechanical response. In particular, much of our work involves studying colloidal suspensions under shear. However, special care must be exercised to usefully implement confocal microscopy in conjunction with our shear apparatus and FMD. First, to study colloidal suspensions, we must choose a suspension that allows for imaging with a confocal microscope and that provides strong stress signals for the FMD. Second, confocal images are always partially distorted due to the confocal's optics. To accurately resolve the sample's microstructure, then, we must understand this distortion and account for it if necessary. Finally, to quantitatively measure the sample microstructure and its dynamics, the confocal image data must be characterized with computational image analysis methods. After discussing the details of these three issues, we illustrate the performance of our shear apparatus in conjunction with confocal microscopy by examining the possibility of wall-slip and non-affine motion in a colloidal suspension, both of which are difficult to check in a conventional rheometer. 
	\subsection{Choosing a suspension for rheoscopic measurements:  index and density matching}

Our shear apparatus is designed to mount onto the stage of a confocal microscope (Zeiss LSM 5 Live inverted confocal microscope). The basic principles of the technique are described in detail elsewhere \cite{Diaspro2002}. In short, a confocal uses a pinhole to control optical sectioning of a fluorescent sample. Focusing light of the appropriate wavelength at one point in the sample causes the sample to fluoresce. The pinhole is placed at a focal point conjugate to the illuminated point, which blocks all out-of-focus light from reaching the detector. Thus, in principle, only a single point is imaged at a time. Scanning this point throughout the sample provides a three-dimensional map of sample regions containing fluorescence dye.  To speed up data acquisition the Zeiss LSM 5 Live instrument scans an entire line at once, instead of a point, but the basic principles remain the same. 

Since confocal microscopy relies on fluorescence rather than transmitted or reflected light, either the particles or the solvent must be dyed.   If the particles are dyed, confocal images show bright spots on a dark background, whereas if the solvent is dyed instead, the spaces occupied by the particles appear dark and the background bright. As long as the refractive index of the solvent closely matches that of the particles, the interior structure of even dense suspensions can be mapped with precision.  However, if the particles' refractive index differs from the solvent, then due to Mie scattering the particles will scatter the illuminating light, in turn creating a turbid sample and severely degrading the image quality. This requirement of matched particle-solvent refractive indices strongly constrains one's choices when selecting a suspension for quantitative experiments using confocal microscopy.  One of the most popular choices is poly-(methyl methacrylate) (PMMA) spheres suspended in a mixture of hydrocarbons such as decalin (decahydronaphthalene) and bromocyclohexane (CXB).  PMMA spheres can be synthesized, stabilized, and dyed using standardized recipes \cite{Antl1986, Campbell2002, Klein2003}.  Hydrocarbon solvent mixtures can be designed to nearly match both the density and the refractive index of the PMMA spheres.  Moreover, PMMA suspensions provide an excellent approximation of hard-sphere interaction dynamics \cite{Royall2013, Campbell2002}.

For stress measurements, however, one also requires the solvent viscosity to be high enough to produce measurable stresses. The calibration curves shown in Section III suggest that, despite the sensitivity of the double cantilever design, it is difficult to access shear stresses below roughly 1 Pa using our apparatus. Thus, shear rates on the order of $10^3$ s$^{-1}$ are needed to produce measurable stresses when the suspension viscosity is similar to that of water, {\it e.g.}, on the order of a few mPa$\cdot$s. At low and intermediate volume fractions many suspensions, including the PMMA standard described above, have viscosities close to this range.  Thus a more viscous solvent-particle system is therefore required if our rheoscopic experiments are to explore broad ranges of shear rates.

Silica particles suspended in glycerol and water provides a useful alternative to PMMA suspensions. A roughly 4:1 mixture of glycerol and water can match the refractive index of silica. Moreover, this mixture has a viscosity of 60 mPa$\, \cdot \,$s, nearly two orders of magnitude larger than that of water and many oils, including decalin, tetralin, and CXB. This silica-based suspension makes it possible to obtain measurable stresses across many orders of magnitude in shear rate using our apparatus.  The silica particles are not density matched with the solvent, however, and thus sedimentation effects need to be considered. The gravitational P\'{e}clet number,
\[  \text{Pe}_g =  \frac{a U}{D}, \]
provides a comparison between sedimentation and diffusion \cite{Russel1989, royall2007nonequilibrium}. Here, the diffusion constant $D$ for a sphere is given by the Stokes-Einstein relation,
\[  D = \frac{kT}{6 \pi a \eta},  \]
where $a$ is the sphere radius and $\eta$ is the viscosity of the surrounding fluid. The sedimentation velocity $U$ for a falling sphere is set by a balance between gravity, buoyancy, and drag:
\[  U = \frac{2 (\rho - \rho_f) a^2 g}{9 \eta},  \]
where $\rho$ and $\rho_f$ are, respectively, the densities of the sphere and the surrounding fluid.  Combining these expressions, we find that 1.0 $\mu$m diameter silica spheres in an index-matching suspension of glycerol and water have a $\text{Pe}_g \approx 0.5$. Although the effect of sedimentation on our system is not negligible, we observe considerable shear-induced viscous resuspension of our colloidal particles \cite{Leighton1986}. Furthermore, since the gravitational P\'{e}clet number ${Pe}_g\propto a^4$, its value quickly reduces to $\sim 0.1$ when particles that have a slightly smaller diameter $2a=0.7$ $\mu$m are used. Thus, the micron-sized silica-based suspensions are fairly well-suited to rheoscopic measurements.

\subsection{Quantifying the confocal's optical response}

\begin{figure}[!t]
\centering
\includegraphics[width=2.8in]{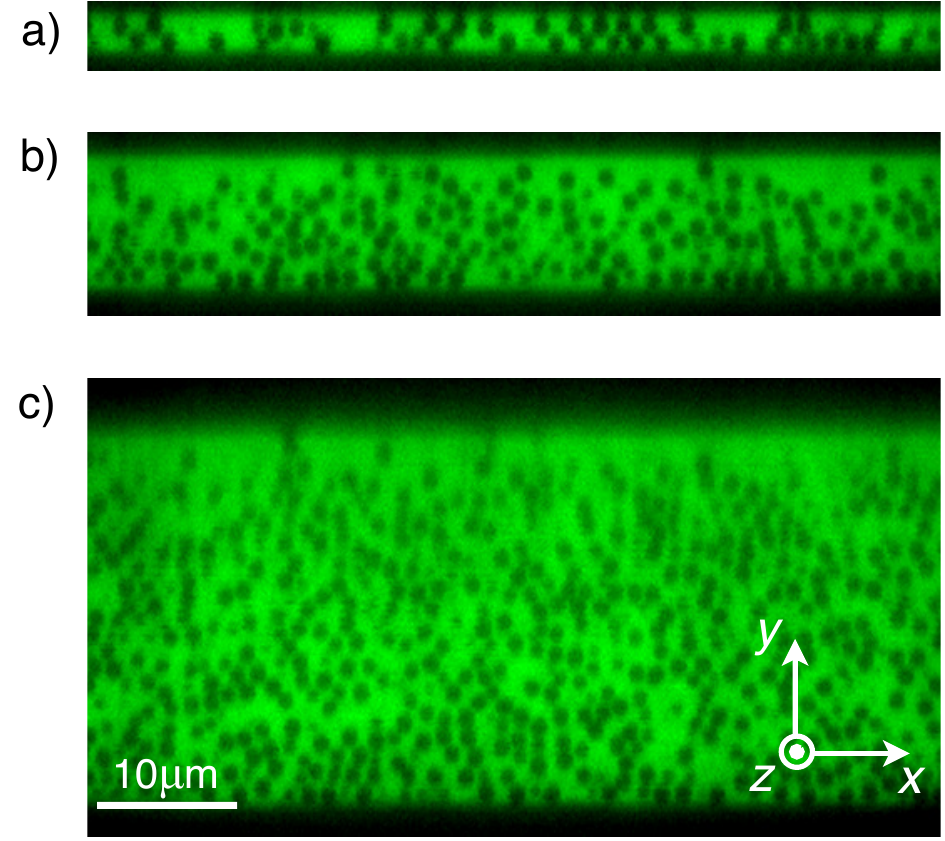} 
\caption{
Images of quiescent colloidal silica suspensions at different gap sizes (a-c). The flow-gradient cross-section is shown (gravity points down in the figure). The dyed solvent appears bright in the image. Since the dye does not penetrate the silica spheres, they appear dark. The glass plate (bottom of cross-sections) and the silicon wafer (top of cross-sections) appear as the dark horizontal regions. Our shear cell allows us to tune the gap size from 3.1 $\mu$m (a) to explore strongly confined systems, to 10.1 or 27.4 $\mu$m (b,c) to explore systems that approach bulk suspensions. However, as is visible in (c), poor image quality and significant aberrations appear when imaging deep into the sample, due to inherent limitations in confocal optical imaging \cite{Hell1992}. 
}
\label{SlicesFig}
\end{figure}

\begin{figure*}[!t]
\centering
\includegraphics[width=5.5in]{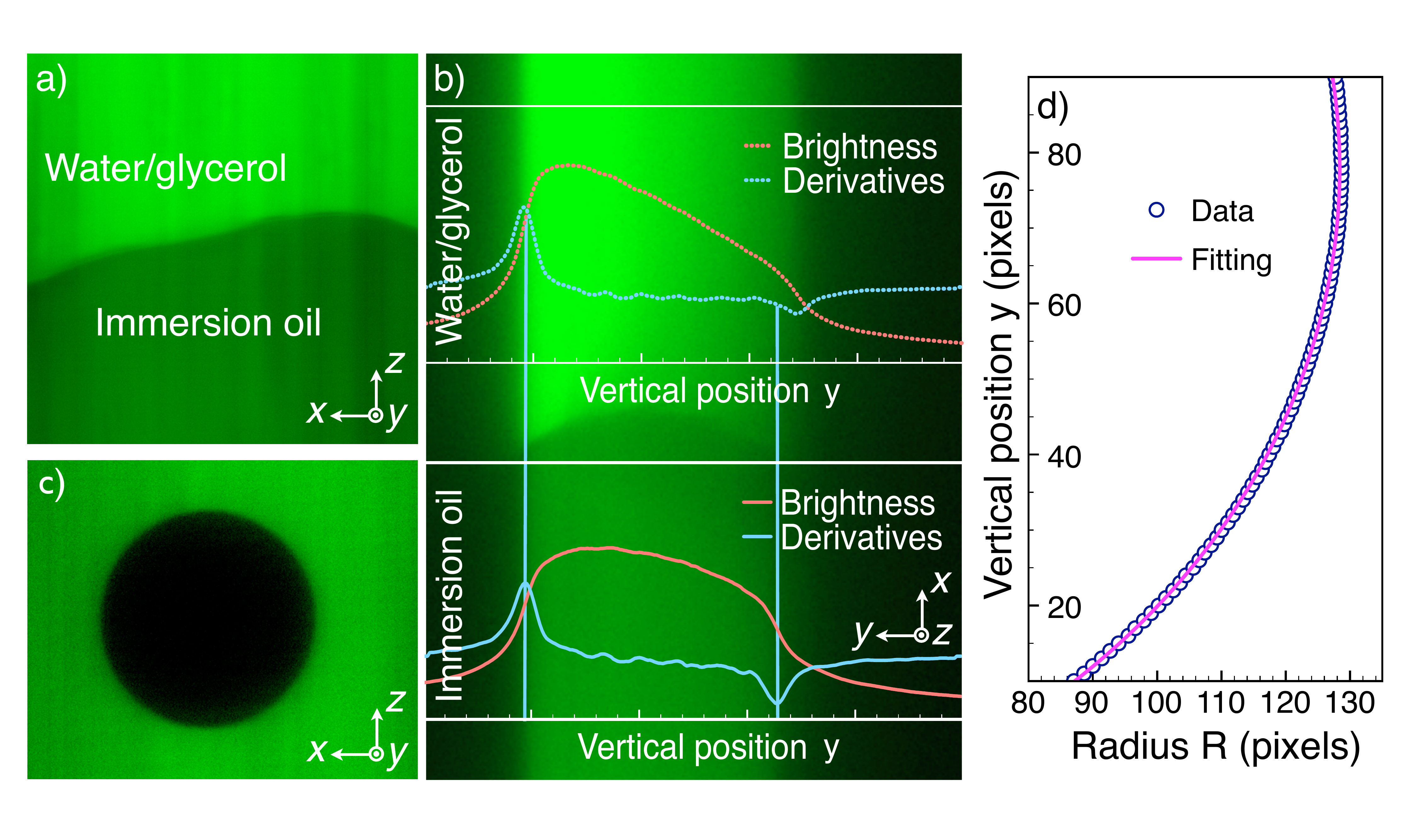} 
\caption{
Measuring the optical distortion along the $y$-direction. (a) $xz$-cross section of the immersion oil: water glycerol image. (b) $xy$-cross section of the same image. Plotted at right is the intensity profile of the image as a function of $y$. In both images, the oil phase is located at the top of the image. At left is the derivative of the above intensity profile. (c) Cross-section of the 30 $\mu$m polystyrene sphere used to find the $y$-voxel size. (d) Blue circles: The apparent sphere radius $R$ in (c) as a function of the apparent $y$-pixel. By fitting the data to an ellipse (red line), we can extract the $y$-voxel height from the known $xz$-pixel values. 
}
\label{Z_PSF1}
\end{figure*}

Before using our confocal microscopes to analyze the suspension's microstructure and rheology, we must first explore the limitations and responses of our confocal microscope. Fig.~\ref{SlicesFig} shows vertical $xy$ slices through three-dimensional images of one of our silica-based suspensions, obtained for gap heights of 3.1 $\mu$m, 10.1 $\mu$m, and 27.4 $\mu$m.  The dark dots are 1$\mu$m silica colloidal spheres and the bright background is the glycerol-water solvent containing fluorescein sodium salt.  The black regions at the top and bottom of each slice represent regions where the focal plane is chosen beyond the upper or lower boundaries of the shear cell, {\it i.e.}, outside the sample volume.  These slices clearly showcase the instrument's ability to create highly confined samples with upper and lower boundaries that can be brought within a few particle diameters of one another. Two imaging artifacts are conspicuous in Fig.~\ref{SlicesFig}, however. First, the spherical particles appear to be stretched along the $y$-direction. Second, the image quality noticeably worsens deeper into the sample. Careful consideration of these effects is important, especially for experiments requiring analysis of scans taken at different depths in the sample. Note that in the microscopy literature what we call the $y$-axis is often referred to as the optical $z$-axis.

The imaging artifacts visible in Fig.~\ref{SlicesFig} arise from two separate issues. First, due to mechanical and software limitations, the actual $y$-direction step size differs from its nominal value. Second, as shown in the previous works \cite{Hell1992, jenkins2008confocal}, there are inherent imaging distortions in confocal microscopy, especially if there is an index mismatch between the sample and the microscope lens. The mismatch between the sample and optics is significantly worse for silica in glycerol and water than for PMMA suspensions. These issues have two major consequences.  First, the center of the microscope's point-spread function, or the confocal's imaging response to a point source, does not move with the lens but instead lags behind. This lagging results in an apparent volumetric pixel (voxel) height that is less than would be expected from the motion of the lens. Second, the confocal's point-spread function in the $y$-direction significantly worsens as images are taken deeper into the sample. Without correcting for these biases, one cannot be confident about particle positions in the $y$-direction.  In addition, the anisotropy in the point spread function and an incorrect $y$-positioning will make a spherical particle appear stretched and can affect featuring of non-spherical particles. 

\begin{figure}[!t]
\centering
\includegraphics[width=3.2in]{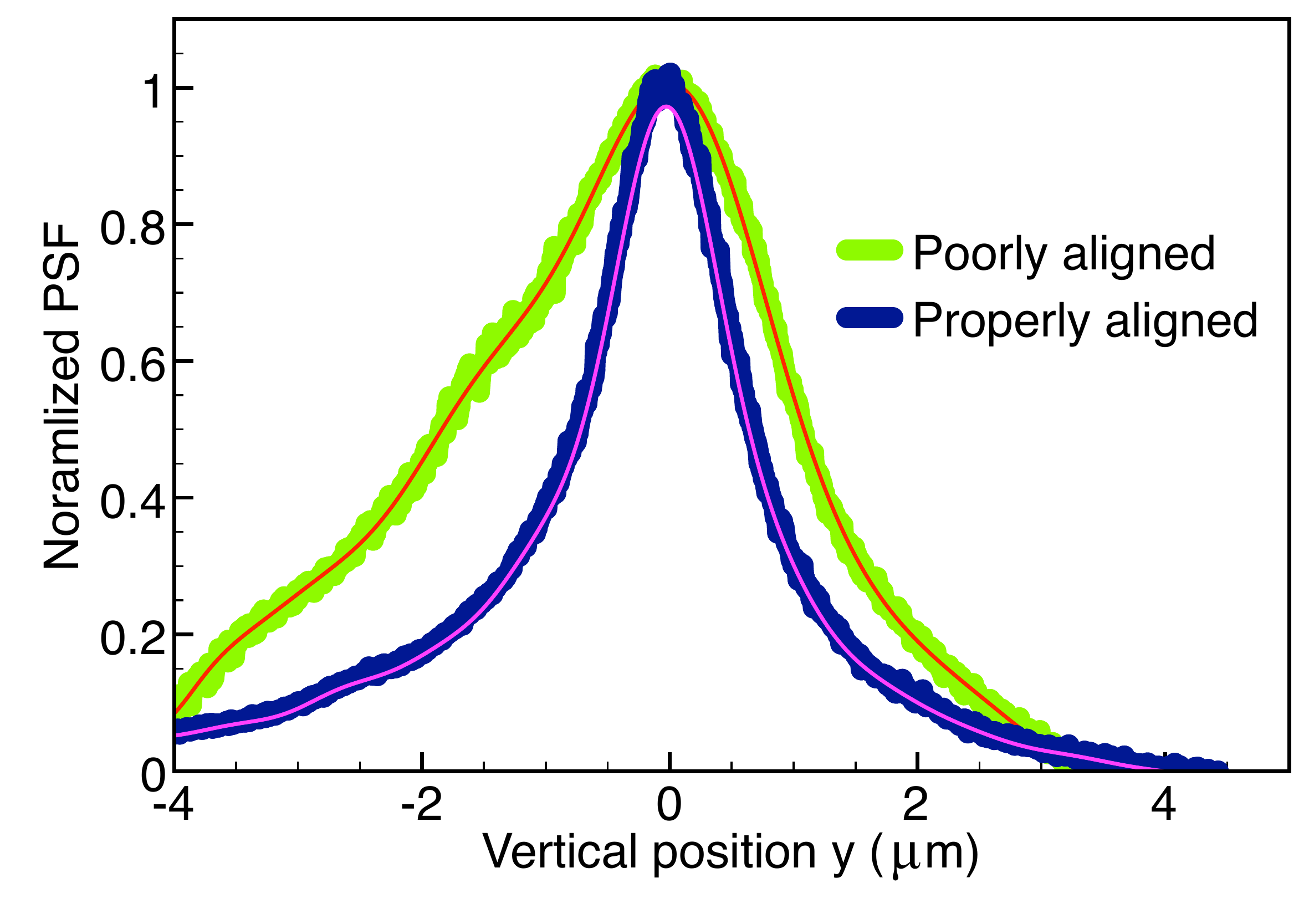} 
\caption{Point-spread function when the confocal is aligned (purple) and when the confocal is misaligned (green). Misalignment of the confocal's internal optics can significantly affect image quality, as visible from the figure. For both curves, the thick lines are clouds of individual data points. The solid thin lines in the foreground are a Legendre polynomial approximation to the measured points, which reduces the noise in the measured PSF. 
} 
\label{PSF_alignment}
\end{figure}

We calibrated the distortion in the $y$-direction using two independent methods. Following Ref.~ \cite{Hell1992}, we first made a sample cell consisting of two pieces of glass in a wedge configuration. One half was filled lengthwise with a mixture of 80:20 water glycerol with fluorescein dye added (used for experiments with silica particles), and the other half with immersion oil with Nile Red dye added. Since the immersion oil has the same index of refraction as the coverslip and the microscope lens, there is minimal optical distortion in the immersion oil region. We took three-dimensional images of the entire gap in each section simultaneously and found the apparent top and bottom of the sample cell in each region. Due to effects of index mismatch, there appears to be a ``jump'' in height across the oil/water interface \cite{Hell1992}, cf.~Fig.~\ref{Z_PSF1}a~\&~b. The difference in heights gives the effect of $y$-distortion due to the index mismatch alone. Moreover, by taking slow image stacks with high slice fidelity, we measured systematic differences between the confocal's scanning stage nominal step size at fast frame rates and the actual $y$-displacement of the lens. By combining the effects of optical distortion and incorrect mechanical step size, we find that when the confocal is programmed to use 0.2 $\mu$m step sizes, the final apparent step size is only 0.166(5) $\mu$m in a water-glycerol solvent, or $\approx$83\% of the input value. Of this, the optical distortion causes a $\approx 8\%$ decrease in the voxel $Y$-size, and the incorrect mechanical step size causes a $\approx 10\%$ decrease in the voxel size. 

Our second method for measuring optical distortion consisted of using large beads. We obtained highly monodisperse 20.85 $\pm$0.04 $\mu$m and 30.39$\pm$0.05  $\mu$m polystyrene spheres from Bangs Laboratories (NIST-traceable grade). For each sphere size, we made a dilute suspension of these spheres in the same 80:20 glycerol-water mixture used for imaging silica spheres, and we mounted this suspension on a microscope sample slide. Since polystyrene is slightly buoyant in the water-glycerol mixture, we placed the coverslip directly onto the sample to keep the polystyrene particles near the interface. We then repeatedly imaged the bottom halves of several spheres. From these images, we used an image edge finding technique to find the radius $\rho$ of the sphere as a function of the apparent $y$ height. Any distortion or incorrect mechanical step sizes will stretch the image along the $y$-direction, making the spherical particle appear as an ellipsoid of revolution. We then fit $\rho(y)$ to an ellipse and extracted the semimajor and semiminor axes of the ellipse. From the ratio of the semimajor to semiminor axes and the known $xz$-pixel value, we can calculate the apparent $y$-pixel value. This method allows us to include effects from both incorrect $y$-positioning of the confocal and from index mismatch of the solvent in one measurement. We repeated this measurement with multiple spheres at each of the two different sphere sizes. From these images, we measure the apparent $y$-pixel ratio to be 0.169(3) $\mu$m with an input 0.2 $\mu$m $y$-step size. Through a simple rescaling of the positions along the gradient direction, these calibrations allow us to accurately measure both the gap size of our shear cell -- crucial for knowing the applied shear rate -- and the microstructure of the colloidal suspension. 

Calibrating the effects of the confocal's point-spread function (PSF) is more difficult. Hell \cite{Hell1992} has shown that the PSF varies with the optical depth into the sample, and that the PSF changes significantly for samples of different refractive indices. 
Rather than measuring the full PSF in a different medium than our sample \cite{Cole2011}, we opted to measure an $xz$-averaged PSF in the same medium as our sample. To do this, we approximated the PSF as translationally invariant near the cover slip. We then imaged a flat interface between a glass slide and the same fluorescein-dyed glycerol-water solution used in our experiments. The resulting image can be expressed as 
	\begin{eqnarray}
			I(x,y,z) &\propto \int_{-\infty}^{\infty} H(y-y') p(x',y',z') dx'dy'dz' \\ \nonumber
			&\propto \int_{-\infty}^{\infty} H(y-y') \tilde{p}(y') dy'
	\end{eqnarray}
where $p(x,y,z)$ is the confocal's point-spread function, and $\tilde{p}$ is the $xz-$averaged PSF. $H(y)$, the Heaviside step function, describes the true intensity profile near the interface. Taking a derivative in $y$ recovers $\tilde{p}(y)$, the $xz$-averaged PSF. This formulation has the additional advantage of averaging over a large field of view to reduce noise in the PSF. To increase our accuracy in the measurement of the PSF, we averaged our measurements over $600$ images. By setting the location of the interface as the value of $y$ such that the measured intensity reaches a fixed fraction of its maximum, we can account for variations in the confocal's $y$-positioning. The resulting $xz$-averaged PSF is shown in Fig.~\ref{PSF_alignment}. Moreover, Fig.~\ref{PSF_alignment} also displays the measured PSF when the confocal's optics are poorly aligned. We find that the PSF is significantly worse in this case, as mentioned in previous work \cite{Cole2011}. Therefore, proper optical alignment is critical for precision measurements. 

	\subsection{Quantifying suspension structure and dynamics with confocal microscopy}

The confocal microscope's three-dimensional scanning ability allows us to check the entire shear zone for dust, bubbles, silicon wafer fragments, and other sample contaminants before beginning an experiment. Under shear, we typically observe uniform behavior throughout the sample but often use only the central region of the cell for quantitative measurements. Our Zeiss LSM 5 Live instrument captures a single 512-pixel row of data at a time. This line, which is oriented in the $x$-direction, is scanned in the $z$-direction to complete a full 512 pixel $\times$ 512 pixel image. For a 100$\times$ microscope objective, this corresponds to a 61.4 $\mu$m $\times$ 61.4 $\mu$m sample slice oriented parallel to the cell boundaries, {\it i.e.}, parallel to the velocity-vorticity plane. At this resolution, a maximum of 60 frames per second can be collected at a fixed height $y$. When the field of view is reduced to 512 pixels $\times$ 128 pixels, the maximum frame rate increases to 216 frames per second. Frames can be collected at different heights to probe three-dimensional structure.  Using standard center-finding algorithms, particle positions can be estimated to sub-pixel accuracy \cite{Crocker1996}. These particles positions can be used to calculate a number of important physical quantities, including flow profiles and correlation functions, as discussed below and elsewhere \cite{Crocker1996, besseling2009quantitative}.

\begin{figure*}[htp]
\centering
\includegraphics[width=0.7\textwidth]{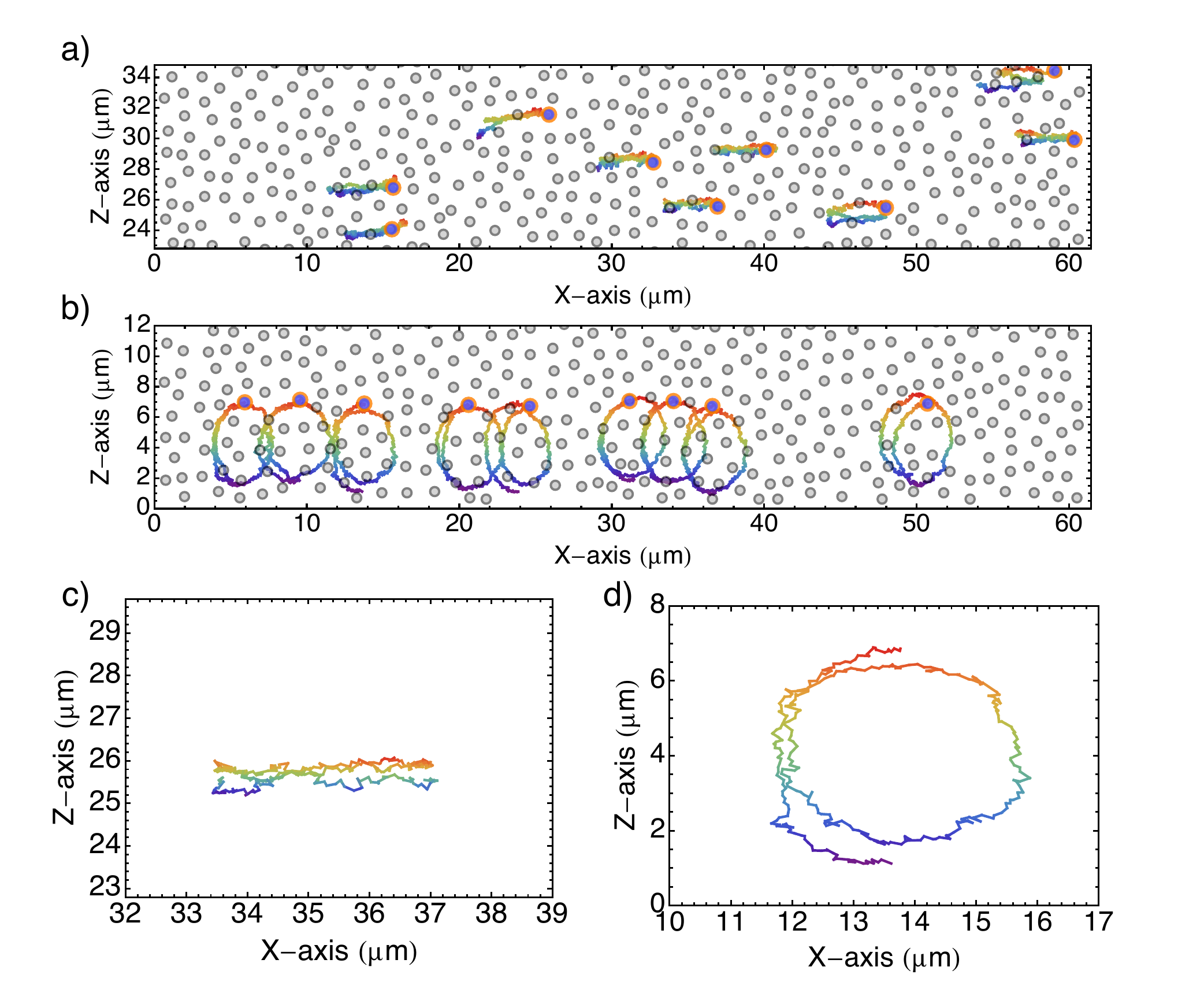}
\caption{
Particle trajectories under linearly polarized (a) and circularly polarized (b) shear, color coded in time over one shear cycle. Particle positions at the start of the cycle are shown in gray filled circles. By tracking individual particles we can not only find the collective motion of the suspension but we can also examine individual particle trajectories. (c,d) A close-up view of the tracked trajectories from (a) and (b), respectively. The trajectories clearly show both the shear-induced motion of the particle as well as a random Brownian component. 
}
\label{tracking_path}
\end{figure*}

Strictly speaking, confocal microscopy does not provide an instantaneous snapshot of particle positions in the imaging plane.  Data is collected from different parts of the sample at slightly different times.  For static samples, this has little quantitative impact.  For suspensions under shear, however, the sample is moving while being scanned and the detector may record distortions associated with these motions.  At the maximum frame rate, the instrument can scan across a 512 pixel $\times$ 512 frame in roughly 0.01 seconds, {\it i.e.}, at a scan speed on the order of 6000 $\mu$m/s.  As a comparison, the maximum velocity $v_{\text{max}}$ of a colloidal particle subjected to sinusoidal shear is given by $v_{\text{max}} = 2 \pi f A$, where $A$ is the displacement amplitude and $f$ is the oscillation frequency. Using the amplitude $A \approx 22$ $\mu$m, we find that $v_{\text{max}}$ is much smaller than 6000 $\mu$m/s for oscillation frequencies on the order of a few Hz or less.  In a sample with a gap height of 5 $\mu$m, for example, shear rates of up to 100 can be reached with these frequencies.  Thus, for slow to moderate shear flows in narrow gap samples, the instrument can be regarded as imaging particle configurations in two dimensions nearly instantaneously.  This range includes, for example, the entire shear thinning regime and much of the Newtonian plateau probed in Ref.~\cite{Cheng2011}.

As an illustration of the usefulness of this approach for direct visualization of suspension dynamics, Fig.~\ref{tracking_path} compares two-dimensional particle motions observed under different shear conditions.  The particles are tracked individually using standard techniques \cite{Crocker1996, besseling2009quantitative}.  Imposing oscillatory shear flow in the $x$-direction, we see that particle trajectories are horizontal on average (Fig.~\ref{tracking_path}a).  A close up view of a single particle trajectory is shown in Fig.\ref{tracking_path}c.  The fluctuations, which are due to Brownian motion, are dominated by the imposed flow when the shear rate is much larger than a characteristic relaxation rate.  A circular shear flow can be generated by imposing oscillatory shear flows along $x$-axis and $z$-axes simultaneously with a phase difference of $\pi/2$.  This generates circular particle trajectories like those shown in Fig.~\ref{tracking_path}b.  A close up view of one of these trajectories is shown in Fig.~\ref{tracking_path}d.  In other situations, real-time access to particle-scale dynamics is useful in other ways.  In crystalline samples, for example, point and line defects can be tracked along with individual particles.  

\begin{figure}[!t]
\centering
\includegraphics[width=3.4in]{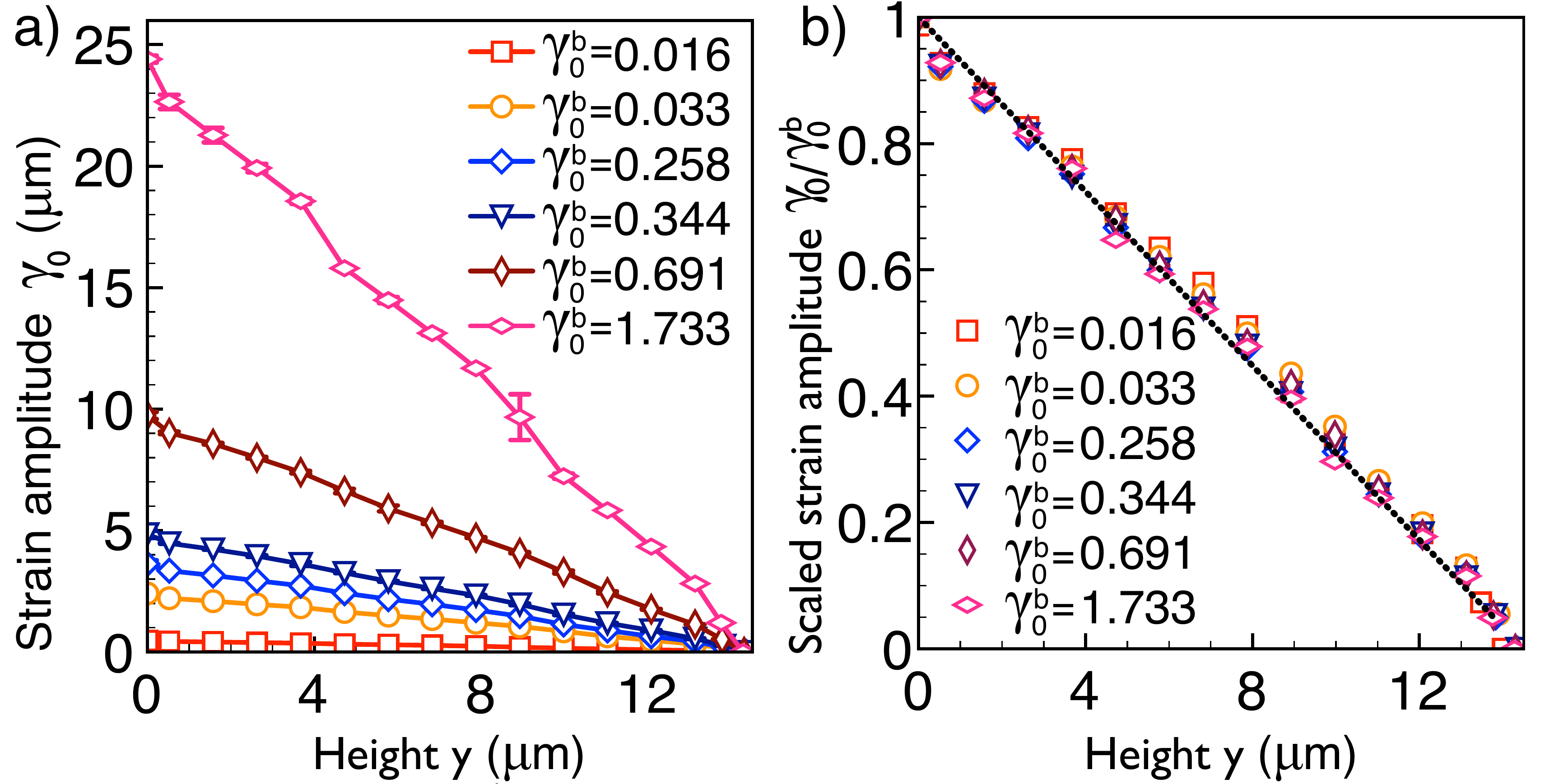}
\caption{ (a) Average particle displacement amplitude as a function of height, for six different strain amplitudes ranging from $\gamma = 0.016$ (red squares) to $\gamma = 1.733$ (magenta diamonds), with an intermediate volume fraction $\phi = 0.48$. All the profiles are linear, and show no shear banding or wall slip. (b) The average particle displacement from (a), normalized to the maximal displacement $\gamma_0^\textrm{bottom}$ which is observed at the bottom of the shear cell. The curves fall on a constant line, as expected for a simple shear flow. } 
\label{shear_profile_combine}
\end{figure}

Due to the symmetries of shear flow between parallel plates, all colloidal particles at the same height move with the same velocity on average. Moreover, these coarse-grained velocities are parallel to the microscope's horizontal imaging plane.  Thus, the bulk velocity field can be extracted by averaging over different in-frame velocity measurements. In practice, for simple oscillatory shear flows, we image a 512 pixel $\times$ 512 pixel window and subdivide it into 31 different overlapping windows, each at 512 $\times $ 32 pixels. Using particle imaging velocimetry (PIV) in each window, we then average over all 31 overlapping windows to find the mean oscillation amplitude and speed of the full 512 $\times$ 512 pixel frame. Repeating this procedure at a series of different heights $y$ builds up an estimate of the flow profile. Fig.~\ref{shear_profile_combine} shows profiles obtained for oscillatory shear at a range of different shear rates $\dot{\gamma}_0$. The profiles collapse when normalized by $\dot{\gamma}_0$ or, equivalently, by the maximum displacement imposed by the piezoelectric device.  The lack of inertial effects in our strongly confined samples (Reynolds number $\textrm{Re} \sim 10^{-6}$) ensures that the instantaneous particle velocities give the instantaneous flow field.  Moreover, our samples are homogeneous apart from slight sedimentation.  We do not observe shear banding in low to intermediate volume fraction suspensions, as one can see in the linear profiles of  Fig.~\ref{shear_profile_combine}.  Together, these properties ensure that the entire sample oscillates in phase to an excellent approximation.  Thus, the amplitudes calculated using PIV are indeed a portrait of the instantaneous bulk flow. 

For sufficiently fast shear, the in-plane structure is distorted by the finite lateral scan-rate.  Since this distortion is linear with $y$ position, simply shifting different lines backwards by different amounts is sufficient to correct the distortion. To obtain an estimate of the instantaneous structure in three dimensions, the vertical scan rate must be taken into account as well.  For a reduced field of view of 512 pixels $\times$ 128 pixels, the maximum scan rate in the $y$-direction is roughly 31 $\mu$m/s.  For a gap size of 5 $\mu$m, then, it takes roughly 0.16 seconds to complete a full stack of images from the bottom boundary of the sample to the top.  Under these conditions, a three-dimensional scan is essentially equivalent to an instantaneous snapshot of the suspension structure only for slow oscillation frequencies of 20-30 mHz or less. For larger frequencies, however, there will be a noticeable mismatch between adjacent image slices, due to the finite vertical scanning speed of the confocal. This distortion can be corrected by linearly shifting the vertical image slices using the PIV data, similar to the method for fixing distortions from the lateral scan rate. A three-dimensional correction scheme is not necessary for flow profiling, since individual frames can be scanned one by one and their contributions to the profile are independent.  The ability to correct distortions in three-dimensions is critical, however, for a statistical analysis of suspension structure.  In particular, a three-dimensional map of particle positions provides access to the pair correlation function, $g(\vec{r})$, the probability of finding a particle at a position $\vec{r}$ relative to another particle's center.  As discussed in the next section, $g(\vec{r})$ provides useful information concerning the relationship between suspension microstructure and bulk rheology.

\section{Applications}


In the previous sections, we have demonstrated that our confocal rheoscope accurately measures the structural and stress responses of a complex fluid in a precisely-controlled shear flow. Similar to other confocal rheometers \cite{Derks2004, Derks2009, Ballesta2008, Besseling2009, Besseling2010, Schmoller2010, Dutta2013, Haw1998a, Haw1998b, Petekidis2002, Cohen2004, Cohen2006, Solomon2006, Besseling2007, Smith2007, Wu2007, Wu2009} this instrument allows us to investigate the interplay between a suspension's structure and its rheology. However, our device also allows for biaxial rheological measurements and the study of confined systems. For example, using this device we have studied shear thinning and thickening of suspensions, the interplay between novel string structures and rheology in confined suspensions, and the rotational as well as translational diffusion of anisotropic particles. Moreover, our device can easily incorporate additional instrumentation or be combined with other imaging techniques such as polarization microscopy to enable and even wider range of studies. 

     \subsection{Shear thinning and thickening of colloidal suspensions}
One generic flow behavior of complex fluids is shear thinning or thickening -- the viscosity either decreasing or increasing, respectively, with increasing shear rate \cite{brader2010nonlinear, mewis2012colloidal}. Colloidal suspensions show both shear thinning and shear thickening behavior. The structural origins of these behaviors have been extensively studied with Stokesian dynamics simulations \cite{foss2000, banchio2003accelerated} and more recently with experiments combining rheometry with light- or neutron-scattering \cite{bender1995optical}. Stokesian dynamics simulations can probe single particle dynamics with unprecedented accuracy in small systems of typically $\sim$10$^3$ particles, given the current limits on computational power. The scattering techniques, on the other hand, measure average behavior of systems consisting of a very large number ($\sim$10$^9$) of particles. The huge gap in system size between these two regimes demands a new bridging technique, which can assess not only the average properties of a suspension, such as its viscosity and normal stress differences, but also the dynamics of individual particles. Using the uniaxial version of this confocal rheoscope we studied the non-Newtonian rheology of suspensions \cite{Cheng2011}. In a typical experiment, our system consisted of 10$^7$ to 10$^8$ particles between two shear plates, which allowed us to obtain accurate average behavior of sheared suspensions. Meanwhile, with the help of fast confocal microscopy, we resolved the motion of single particles over a long period of time. Statistical errors were reduced significantly by averaging over the $\sim$10$^4$ particles within the imaging field of view.

Using our confocal rheoscope, we investigated the configuration of particles under oscillatory shear in the shear thinning regime (around $\textrm{Pe}_s \equiv \dot{\gamma} d^2 / D_s \sim$ 1, where $\dot{\gamma}$ is the shear rate, $d=2a$ the particle diameter, and $D_s$ is the particle self-diffusion constant) \cite{cheng2012assembly}. Using the real-time particle positions measured by our confocal microscope, we examined the pair correlation function of particles in the plane of shear, $g(x,y)$, under a full cycle of oscillatory shear. The fore-and-aft asymmetry developed near the maximum shear rate within the cycle is consistent with previous theories and experiments for suspensions under steady shear flow \cite{brady1993rheological, gao2010direct, vermant2005flow, zia2010single}. The temporal resolution of high-speed confocal microscopy allows us to track the configurations of particles \textit{in situ}. We can further relate the pair correlation function $g(\vec{r})$ to the Brownian stress by\cite{brady1993rheological}:
\begin{eqnarray}
  \tau_B \approx -n^2 k_{\rm B}T a \int\limits_{r = 2a} \hat{r} \hat{r} g(\vec{r}) {\rm d}S
  \label{eq:Brownian_stress} 
\end{eqnarray}
where $n$ is the number density of particles, $k_{\rm B}T$ is the thermal energy, $a$ is the particle radius, and $\hat{r}$ is the unit vector in $r$-direction. Due to resolution limitations in the experiments the integral is evaluated over a small range of radii centered at $r = 2a$ and a prefactor is introduced to compensate for the adjusted integral range in the stress calculation. The total stress can then be calculated by adding the hydrodynamic stress to the Brownian stress \cite{Cheng2011}. Comparison between the total calculated stress determined from the particle positions and the direct measurement with our Force Measurement Device show quantitative agreement.

Furthermore, we also probed the linear viscoelastic behavior of the Brownian stress under oscillatory shear. We fit the real and imaginary parts of the complex Brownian viscosity, as calculated from Brownian stress (Eq.~\ref{eq:Brownian_stress}), to a linear viscoelastic response \cite{Cheng2011}. Fitting these curves indicates a relaxation time on order of 30 s, consistent with the Brownian diffusion time-scale of a dilute suspension. 

More recently, we also investigated the large amplitude oscillatory shear response of suspensions\cite{lin2013far}. By varying the shear amplitude and frequency separately, large amplitude oscillatory shear is able to disentangle the underlying dynamics that are usually convolved in far-from-equilibrium systems. In contrast to the response in the linear regime, the suspension structure response under large amplitude oscillatory shear demonstrates a nonlinear saturation that arises from shear-induced advection. We also showed that in spite of the distinct underlying mechanisms giving rise to the linear and nonlinear responses, all data can be scaled onto a master curve that links small-amplitude oscillatory shear with continuous shear \cite{lin2013far}.

Finally, we also studied the configurations of particles for large Pe $\gtrsim$ 4,000 in the weak shear thickening regime. Reconstructed images show that the suspended particles form a clustered structure \cite{Cheng2011}. This structure is preferentially aligned along the compression axis of the shear. The result is consistent with the prediction on the emergence of hydro-clusters during shear thickening  \cite{wagner2009shear, lee2003dynamic, maranzano2001effects}. Future investigations of hydro-cluster dynamics should elucidate the mechanisms linking hydro-cluster formation and interactions with shear thickening. 

     \subsection{String structure of confined colloidal suspensions}
     
Our confocal rheoscope also allows us to investigate suspension structure and rheology in a confined geometry \cite{cheng2012assembly}. When we shear a confined suspension, one with less than 10 layers of particles, we observe a strong vorticity-aligned string structure at intermediate volume fractions 0.34 to 0.4 with $80 \lesssim \textrm{Pe} \lesssim 4,000$. This vorticity-aligned string structure contrasts with previous simulation studies, where \textit{flow}-aligned string phase have been observed instead. The vorticity-aligned string structure can be attributed to the combination of strong interparticle hydrodynamic couplings and the interlayer momentum exchange in the confined sample \cite{cheng2012assembly, zurita2012layering}. 

Employing our biaxial shear cell, we examined this far-from-equilibrium string structure under two-dimensional oscillatory shear\cite{lin2014}. Using biaxial shear we have unprecedented control over the suspension behavior. For example, we imposed two orthogonal shear flows at the same frequency with different phases $\delta$. If the shear flows are in-phase ($\delta = 0$) the resulting shear flow is a uniaxial oscillatory shear and the string structure is very pronounced. If the shear flows are out-of-phase ($\delta = \pi/2$) the resulting shear flow is a circularly polarized shear. We find that the particle alignment into strings decreases gradually with increasing $\delta$ and eventually becomes isotropic when $\delta=\pi/2$. 

We also investigated the effect of the particle string configuration on the suspension rheology, using our biaxial FMD\cite{lin2014}. Surprisingly, as the suspension morphology progresses from string structures to an isotropic state, we see no corresponding change in the suspension rheology. To clarify the lack of the string structure's rheological signature, we performed an ``oscillatory superposition spectroscopy'' measurement on the suspension. While the suspension was under a uniaxial oscillatory shear flow, with particles assembled into strings, we applied a second, high-frequency oscillatory shear flow. From this second flow, we probed the stress response of the sample both parallel and orthogonal to the primary flow. We found that the stress response is isotropic at the frequencies probed, despite the highly anisotropic suspension string structure. Moreover, the flow behavior is Newtonian both along and orthogonal to the applied flow\cite{lin2014}. These observations highlight our device's capability to investigate novel structures of highly confined samples and test their anisotropic rheological properties.

	\subsection{Particle diffusion under shear}
Our shear apparatus can also be used to study particle dynamics including the translational and rotational diffusion of colloidal particles under shear. Due to Taylor dispersion \cite{Taylor1953, beckman1994recognition} particles undergo faster translational diffusion along the flow direction during shear. By using our confocal microscope in conjunction with our shear apparatus, we were able to measure two additional types of enhanced diffusion under shear \cite{leahy2013enhancing}. Whereas enhanced translational diffusion relies on Brownian motion and an inhomogeneous flow field, enhanced rotational diffusion relies of the inhomogeneous orientation flow field due to the Jeffrey orbits exhibited by the particles. This enhancement of rotational diffusion may allow for interesting self-assembly or rheological applications. Similarly, by looking at dense suspensions of spherical particles, we can use our confocal rheoscope to measure that colloidal particles' diffusion is also enhanced \textit{perpendicular} to the flow direction \cite{cheng2012assembly}. This enhancement arises from hydrodynamic interactions between particles giving diffusive behavior.



	\subsection{Other applications}
While we have only discussed applications of our shear apparatus for simple colloidal liquids, this device is designed to be easily customized for additional applications. For example, with our current design, we can use our shear apparatus in conjunction with a holographic optical tweezers to locally control suspension structure \cite{vossen2004optical}. In principle, this allows us to manipulate the suspension structure down to the single particle scale in a sheared sample. Our three-axis piezoelectric stage additionally allows us to investigate compressional or extensional flows in complex fluids \cite{doyle1998relaxation, gupta2000extensional}, simply by taking advantage of the y-positioning capabilities of the piezo to move the plates perpendicular to the sample boundaries. Moreover, by mis-aligning the top and bottom plates, we can use our shear apparatus to investigate shear or compressional lubrication flows in complex fluids. Our biaxial FMD also allows for probing the anisotropic viscosities of a variety of other complex fluids ranging from colloidal crystals to collagen fiber networks. Moreover, we can access a vast array of additional experimental approaches with only minor modifications to our shear or imaging apparatus. By substituting a transparent cover slip for the opaque FMD, we can use cross-polarized microscopy to investigate the structural dynamics of sheared liquid crystals. Alternatively, we can introduce polarizers and use epifluorescence to conduct fluorescence confocal polarizing microscopy (FCPM) \cite{smalyukh2001three} and measure the 3D director orientation of sheared liquid crystals. To investigate the effects of different boundary structures on the rheology of confined suspensions, we can replace the silicon wafers and glass cover slips with patterned surfaces \cite{jiang2004large, aizenberg2000patterned, lin2000entropically}. Our shear apparatus can be easily modified for use in conjunction with Dynamic Light Scattering, by using a FMD with a small window for a laser beam to pass through. Moreover, due to our modular design of our FMD, in the future it will be easy to substitute an FMD that additionally measures normal stresses. The capabilities of our shear apparatus and its flexibility for a wide range of applications promise that similar confocal rheoscopes will find important uses in future experiments in rheology and soft matter. 


\section*{Acknowledgements}
The authors would like to acknowledge J. Mergo, T. Beatus, and Y.-W. Lin for technical help and useful discussions on apparatus design. Because this technique took years to develop many individuals contributed to its design. The original prototype was developed by I.C. in collaboration with T. G. Mason and D. A. Weitz. The current version of the multiaxis piezo actuator and uniaxial FMD were developed by J.H.M., J.N.I., and I.C., the biaxial shear protocols and FMD were developed by N.L., J.N.I, and I.C. In addition, the arduous work of developing mechanical and optical calibrations, method development, and developing operating procedures were worked on by N.L., X.C., B.L., and J.H.M.. This publication was based on work supported by Award No. KUS-C1-018-02, made by King Abdullah University of Science and Technology (KAUST), the National Science Foundation under Grant No. (DMR 1056662); the US Department of Energy, Office of Basic Energy Sciences, Division of Materials Sciences and Engineering under Award No. ER46517, and in part under National Science Foundation CBET-PMP Award No. 1232666. J.H.M. was funded in part by Colby College,  B.L. acknowledges the DoD, Air Force Office of Scientific Research, National Defense Science and Engineering Graduate (NDSEG) Fellowship 32 CFR 168a. J.N.I was supported by the Department of Energy, Office of Basic Energy Sciences, Division of Materials Sciences and Engineering, under Award DE-FG02-87ER-45331, and by the National Science Foundation NSF grant CHE-1059108.

\newpage


%

\end{document}